\documentclass[twocolumn,english,aps,prl,superscriptaddress]{revtex4-1}

\usepackage{mathpazo}

\usepackage[T1]{fontenc}
\usepackage[latin9]{inputenc}
\usepackage{color}
\usepackage{babel}
\usepackage{amsmath}
\usepackage{amssymb}
\usepackage{graphicx}
\usepackage[citecolor=blue,filecolor=black,linkcolor=red,urlcolor=blue,breaklinks=true,colorlinks=true]{hyperref}

\begin{document}

\title{Quasi-one-dimensional approximation for Bose-Einstein condensates
transversely trapped by a funnel potential}

\author{Mateus C. P. dos Santos}
\address{Instituto de Física, Universidade Federal de Goiás, 74.690-970, Goiânia,
Goiás, Brazil}

\author{Boris A. Malomed}
\address{Department of Physical Electronics, School of Electrical Engineering,
Faculty of Engineering, Tel Aviv University, Tel Aviv 69978, Israel}
\address{ITMO University, St. Petersburg 197101, Russia}

\author{Wesley B. Cardoso}
\address{Instituto de Física, Universidade Federal de Goiás, 74.690-970, Goiânia,
Goiás, Brazil}
\email{wesleybcardoso@ufg.br (WB Cardoso)}

\begin{abstract}
Starting from the standard three-dimensional (3D) Gross-Pitaevskii
equation (GPE) and using a variational approximation, we derive an
effective one-dimensional nonpolynomial Schrödinger equation (1D-NPSE)
governing the axial dynamics of atomic Bose-Einstein condensates (BECs)
under the action of a singular but physically relevant \textit{funnel-shaped}
transverse trap, i.e., an attractive 2D potential $\sim-1/r$ (where
$r$ is the radial coordinate in the transverse plane), in combination
with the repulsive self-interaction. Wave functions of the trapped
BEC are regular, in spite of the potential's singularity. The model
applies to a condensate of particles (small molecules) carrying a
permanent electric dipole moment in the field of a uniformly charged
axial thread, as well as to a quantum gas of magnetic atoms pulled
by an axial electric current. By means of numerical simulations, we
verify that the effective 1D-NPSE provides accurate static and dynamical
results, in comparison to the full 3D GPE, for both repulsive and
attractive signs of the intrinsic nonlinearity. 
\end{abstract}

\maketitle

\section{Introduction}

The observation of the Bose-Einstein condensation (BEC) in alkali-metal
atomic gas at ultralow temperatures was reported 70 years after it
was predicted \cite{Anderson_SCI95,Davis_PRL95,Bradley.78.985}.
Currently, BECs are produced and manipulated by various research groups,
for various kinds of atomic species \cite{Gries.94.160401,Fried_PRL98,Aika.108.210401},
and are also observed in molecular gases \cite{Jochim_SCI03,Zwierlein.91.250401}.
These setting give rise to countless effects, such as the formation
of matter-wave dark \cite{Burger_PRL99,Denschlag_SCI00} and bright
\cite{Khaykovich_SCI02} solitons, propagation of matter-wave soliton
trains \cite{Strecker_NAT02}, creation of vortex states \cite{Matthews_PRL99,Madison_PRL00},
prediction of stable vortex solitons in various forms \cite{Malomed_arxiv19},
observation of Feshbach resonances \cite{Inouye_NAT98}, etc. A recent
addition to this list is the prediction \cite{Petrov_PRL15,Petrov_PRL16}
and experimental creation \cite{Cabrera_SCI18,Cheiney_PRL18,Semeghini_PRL18,Ferioli_PRL19,D'Errico_ArXiv19}
of multidimensional ``quantum droplets'', in which the matter-wave
collapse is arrested by the Lee-Huang-Yang (LHY) effect, produced
by fluctuational corrections to the mean-field dynamics.

The study of static and dynamical properties of dilute BECs at zero
temperature is accurately modeled by the time-dependent three-dimensional
(3D) Gross-Pitaevskii equation (GPE) \cite{Pitaevskii_03,Pethick_08}
(if necessary, complemented by the above-mentioned LHY terms \cite{Petrov_PRL15,Petrov_PRL16}).
A relevant problem is to reduce the full three-dimensional GPE to
lower-dimensional equations, when the reduction is imposed by an external
potential which tightly confines BEC in one or two transverse directions.
Several approximations for the dimensional reductions 3D $\rightarrow$
1D and 3D $\rightarrow$ 2D were elaborated, for both self-repulsive
and attractive signs of the nonlinearity in GPE \cite{Muryshev_PRL02,Salasnich_PRA02,Salasnich_PRA02-2,Massignan_PRA03,Kamchatnov_PRA04,Carr_PRL04,Zhang_PRA05,Salasnich_PRA06,Matuszewski_PRA06,Wei_CPL07,Salasnich_PRA07,Salasnich_PRA07-2,Mateo_PRA07,Maluckov_PRA08,Salasnich_PRA08,Salasnich_JPA09,Li_CTP09,Buitrago_JPB09,Adhikari_NJP09,Salasnich_PRA09,Munoz-Mateo_AP09,Young_PRA10,Cardoso_PRE11,Mateo_PRA11,Edwards_PRE12,Diaz_JPB12,Salasnich_PRA13,Salasnich_PRA14,Chiquillo_JPA15,Calixto1}.
In particular, the use of the variational approximation (VA) for the
transverse profile of the wave function, presented in Refs. \cite{Salasnich_PRA02,Salasnich_PRA02-2},
helps to derive an effective 1D nonpolynomial Schrödinger equation
(1D-NPSE), which accurately models the axial dynamics of the cigar-shaped
BECs \cite{Cuevas_NJP13}. This method can also be applied for the
derivation of effective 2D-NPSEs, when BEC is strongly confined in
the axial direction \cite{Salasnich_PRA02,Salasnich_PRA09}. This
method can be used to analyze the evolution of elongated BECs with
\cite{Massignan_PRA03} or without \cite{Kamchatnov_PRA04} the
assumption that the wave function slowly varies along the axial direction,
dynamics of spin-1 condensates \cite{Zhang_PRA05}, and binary self-attractive
BECs \cite{Salasnich_PRA06}, as well as the behavior of BEC in a
nearly-1D cigar-shaped trap with the transverse confining frequency
periodically modulated along the axial direction \cite{DeNicola_PLA06,Salasnich_PRA07,Salasnich_PRA07-2,Maluckov_PRA08},
effects of embedded axial vorticity in the elongated BECs \cite{Salasnich_PRA08},
matter waves under anisotropic transverse confinement \cite{Salasnich_JPA09},
BECs in \textit{funnel}- \cite{Li_CTP09} and tube-shaped \cite{Calixto1}
potentials, mean-field equations for cigar- and disc-shaped Bose-Fermi
mixtures and fermion superfluids \cite{Buitrago_JPB09,Adhikari_NJP09,Diaz_JPB12},
solitons and solitary vortices in BECs \cite{Salasnich_PRA09}, BECs
in mixed dimensions \cite{Young_PRA10}, dilute bosonic gases with
intrinsic two- and three-body interactions \cite{Cardoso_PRE11},
BECs confined in ring-shaped potentials \cite{Edwards_PRE12}, BECs
with spin-orbit and Rabi couplings \cite{Salasnich_PRA13,Salasnich_PRA14,Chiquillo_JPA15,Li_NJP17,Li_PRA18},
etc.

In this work, we consider BEC confined by a \textit{funnel-shaped}
potential acting in the transverse plane $\left(x,y\right)$, i.e.,
the 2D attractive potential 
\begin{equation}
V_{\mathrm{transverse}}(r)=-\frac{\varepsilon^{3}}{2r},\label{transverse}
\end{equation}
with $\varepsilon>0$ and transverse radial coordinate $r\equiv\sqrt{x^{2}+y^{2}}$.
As suggested by Ref. \cite{Sakaguchi_PRA11}, this potential may
be applied by an axially oriented wire, uniformly charged with density
$\sigma$, to small molecules carrying a permanent dipole moment $\mathbf{d}$,
which is oriented along the local electric field, with strength of
the radial electric field $\mathbf{E}=2\sigma\boldsymbol{r}/r^{2}$
(the interaction of a neutral polarizable atom with a uniformly charged
wire was considered in Ref. \cite{Denschlag_EPL97}). Then, indeed,
the potential pulling the dipoles to the axis is $V_{\mathrm{transverse}}(r)=-\mathbf{d}\cdot\mathbf{E}\equiv-2\sigma d/r$,
which is tantamount to potential (\ref{transverse}). A still more
realistic possibility is provided by a quantum gas of atoms carrying
permanent magnetic moments (such as chromium, dysprosium, or erbium
\cite{Gries.94.160401,Ferrier-Barbut_PRL16,Schmitt_NAT16,Chomaz_PRX16}),
pulled in the axial direction by an electric-current jet (which may
be an electronic beam) \cite{Sakaguchi_PRA11}.

With the help of VA, we derive an effective 1D equation describing
the axial dynamics in this setting. A similar configuration was considered
in Ref. \cite{Li_CTP09}. However, unlike that work, which used VA
based on the Gaussian \emph{ansatz}, we employ one with an exponential
profile. This difference is more than a formal modification, as the
exponential approximation agrees with the specific asymptotic form
of the wave function at $r\rightarrow0$, imposed by the singularity
of the funnel trap (see Eq. ((\ref{rho3})) below), while the Gaussian
\emph{ansatz} contradicts it. As a result, full 3D numerical simulations
demonstrate that we obtain a more accurate model equation for the
BEC confined by the funnel-shaped potential (in fact, no solutions
of the 1D equation, nor the asymptotic form of the 3D wave function,
were considered in Ref. \cite{Li_CTP09}). Limit cases, pertaining
to weak- and strong-coupling regimes, are also considered, as well
as results provided by the effective equation derived in Ref. \cite{Li_CTP09},
for the comparison's sake. Further, we demonstrate that the effective
NPSE obtained here also provides an accurate approximation for the
transverse width of the confined BEC.

The rest of the paper is organized as follows. In the next section,
starting from the 3D GPE and using the VA, we derive an effective
NPSE governing the axial dynamics in the funnel-shaped configuration.
Limit cases of the NPSE, and a similar effective 1D equation which
was derived, in a more formal way, in Ref. \cite{Li_CTP09} are addressed
in the same section. Systematic numerical results, for both the static
ground states (GSs) and dynamical regimes, are reported in Sec. \ref{NR}.
The paper is concluded by Sec. \ref{Conclusion}.

\begin{figure}[tb]
\centering \includegraphics[width=0.9\columnwidth]{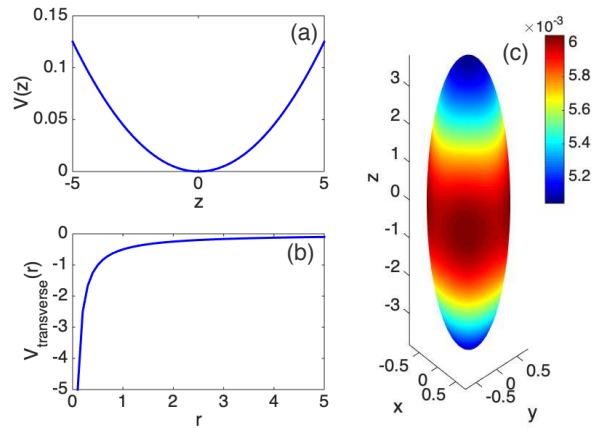} \caption{The confining potential, given by Eqs. (\ref{TPOT}) and (\ref{eq:POT1}),
in the axial (a) and radial (b) directions. (c) A schematic shape
of the density in the respective 3D ground state.}
\label{F0} 
\end{figure}

\begin{figure*}[t]
\centering \includegraphics[width=0.8\textwidth]{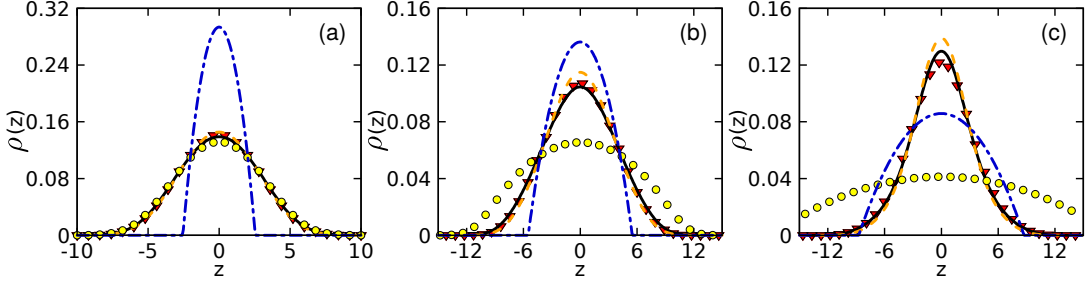} \caption{Normalized axial density profiles $\rho(z)$ for a repulsive BEC ($g>0$),
under the combined action of the funnel trap (\ref{transverse}) with
$\varepsilon=1$ and axial potential (\ref{eq:POT1}) with $\lambda=0.1$,
for (a) $g=1$, (b) $g=10$, and (c) $g=40$. The density profiles
were numerically obtained as GS (ground-state) solutions of the full
3D GPE (\ref{3DGPE_scaled}) (red triangles), 1D-NPSE (\ref{eq:EFF})
(the black solid line black), 1D YXWLH equation (\ref{YXWLH}) (the
orange dashed line), cubic 1D-NLSE (\ref{eq:W}) (yellow circles),
and the TF state (\ref{eq:TF}) (the blue dashed-dotted line).}
\label{F1} 
\end{figure*}

\section{The derivation of the effective nonpolynomial Schrödinger equation
\label{DENPSE}}

A monatomic BEC at zero temperature is modeled by the commonly known
3D GPE, 
\begin{equation}
i\hbar\frac{\partial\psi}{\partial t}=-\frac{\hbar^{2}}{2m}\mathbf{\nabla}^{2}\psi+V(r,z)\psi+\mathcal{N}U|\psi|^{2}\psi,\label{3DGPE}
\end{equation}
where $\psi(\mathbf{r},t)$ is the mean-field wave function of the
condensate, potential 
\begin{equation}
V(r,z)\equiv-\frac{\varepsilon^{3}}{2r}+V(z)\label{TPOT}
\end{equation}
includes the funnel trapping term (\ref{transverse}) and a generic
(nonsingular) axial one, $V(z)$, $U\equiv4\pi\hbar^{2}a_{s}/m$ is
the nonlinearity strength, with $a_{s}>0$ being the atomic \textit{s}-wave
scattering length of repulsive interatomic interactions, and $\mathcal{N}$
is the number of condensed atoms. By means of rescaling $t\rightarrow\omega_{r}t$,
$(x,y,z)\rightarrow(x,y,z)/a_{r}$, $\psi\rightarrow\psi a_{r}^{3/2}$,
with $a_{r}$ being a characteristic length scale in the transverse
direction, Eq. (\ref{3DGPE}) is cast in the normalized form, 
\begin{equation}
i\frac{\partial\psi}{\partial t}=-\frac{1}{2}\mathbf{\nabla}^{2}\psi+V\left(r,z\right)\psi+2\pi g|\psi|^{2}\psi,\label{3DGPE_scaled}
\end{equation}
with $g\equiv2\mathcal{N}a_{s}/a_{r}$.

As concerns the above-mentioned physical interpretation of the model
in terms of the condensate of small molecules carrying the permanent
electric moment, which are pulled to the uniformly charged axial wire,
the respective GPE must take into account the long-range dipole-dipole
interaction between the particles. As shown in Ref. \cite{Sakaguchi_PRA11},
this can be done by introducing a mean-field electrostatic field,
which is induced by the effective charge density in the dipolar gas
(as per the respective Poisson equation), $\mathbf{E}_{d}=-4\pi\left\vert \psi\left(x,y,z\right)\right\vert ^{2}\mathbf{d}$.
This, in turn, gives rise to effective renormalization of the scattering
length, which is written in terms of the unscaled notation, $a_{s}\rightarrow a_{s}+md^{2}/\hbar^{2}$.
Similarly, the mean-field approximation makes it possible to take
into account dipole-dipole interactions between magnetic atoms.

A schematic representation of the trapping potential (\ref{TPOT}),
and expected distribution of the BEC density in the corresponding
GS is displayed in Fig. \ref{F0}. Here, the axial term of the potential,
$V(z)$, is assumed to take the harmonic-oscillator (HO)\ form, see
Eq. (\ref{eq:POT1}) below.

In spite of the presence of the singular potential (\ref{transverse}),
Eq. (\ref{3DGPE_scaled}) gives rise to a GS\ wave function with
chemical potential $\mu$ and a regular expansion at $r\rightarrow0$:
\begin{eqnarray}
\psi\left(r,z.t\right) & = & e^{-i\mu t}\left[\left(1-\varepsilon^{3}r\right)\psi_{0}(z)+\frac{1}{2}\psi_{2}(z)r^{2}+...\right],\label{rho3}\\
\psi_{2}(z) & = & \left(\frac{1}{2}\varepsilon^{6}-\mu+V(z)\right)\psi_{0}(z)+2\pi g\psi_{0}^{3}(z)-\frac{1}{2}\psi_{0}^{\prime\prime}(z)\nonumber 
\end{eqnarray}
{[}function $\psi_{0}(z)$ can be uniquely determined only from the
global solution of Eq. (\ref{3DGPE_scaled}), rather than solely from
the expansion at small $r${]}. An essential difference of Eq. (\ref{rho3})
from a similar expansion constructed with a nonsingular potential
is that, in the latter case, the expansion does not contain a term
$\sim r$. On the other hand, it is relevant to stress that the GS
\emph{does not exist} in the presence of a more singular radial potential,
\textit{viz}., $V_{\mathrm{transverse}}(r)=-Cr^{-2},$ with $C>0$.
In that case, one can formally find a wave function with a singularity
at $r\rightarrow0$, $\psi\sim r^{-1}$; however, this solution is
unphysical, as its norm diverges \cite{Sakaguchi_PRA11}.

The Lagrangian density, corresponding to Eq. (\ref{3DGPE_scaled})
with confining potential (\ref{TPOT}), is 
\begin{eqnarray}
\mathcal{L} & = & \frac{i}{2}\left(\psi^{\ast}\frac{\partial\psi}{\partial t}-\psi\frac{\partial\psi^{\ast}}{\partial t}\right)-\frac{1}{2}|\nabla\psi|^{2}\nonumber \\
 &  & -\left(V(z)-\frac{\varepsilon^{3}}{2r}\right)|\psi|^{2}-\pi g|\psi|^{4}.\label{eq:ACT}
\end{eqnarray}
Our goal is to reduce the 3D model to an appropriate 1D approximation.
To this end, we consider the following factorized\emph{\ ansatz}:
\begin{equation}
\psi(r,z,t)=\exp\left(-\frac{r}{2\eta^{2}}\right)\frac{f(z,t)}{\sqrt{2\pi}\eta^{2}},\label{eq:ANZ}
\end{equation}
were $f(z,t)$ is an axial wave function, and $\eta=\eta(z,t)$ is
the transverse width, with the normalization condition 
\begin{equation}
\int_{-\infty}^{+\infty}|f\left(z\right)|{^{2}dz}=1,\label{N}
\end{equation}
which provides unitary normalization of the full 3D wave function
(\ref{eq:ACT}). Note that the expansion of \emph{ansatz} (\ref{eq:ANZ})
at $r\rightarrow0$ agrees with the asymptotic form given by Eq. (\ref{rho3}).

The effective 1D-NPSE is produced by inserting \emph{ansatz} (\ref{eq:ANZ})
into Lagrangian density (\ref{eq:ACT}) and performing integration
in the transverse plane, it being relevant to stress that the 2D integral
with singular term $\sim1/r$ in Eq. (\ref{eq:ACT}) converges. This
procedure leads us to the following effective 1D Lagrangian, 
\begin{eqnarray}
L_{\mathrm{1D}} & = & \int_{-\infty}^{+\infty}\left[\frac{i}{2}\left(f^{\ast}\frac{\partial f}{\partial t}+\mathrm{c.c.}\right)-\frac{1}{2}\left\vert \frac{\partial f}{\partial z}\right\vert ^{2}\right.\nonumber \\
 & - & \left.\left[V(z)+\frac{1}{2\eta^{2}}\left(\frac{1}{4\eta^{2}}-\varepsilon^{3}\right)\right]|f|^{2}-\frac{g}{8}\frac{|f|^{4}}{\eta^{4}}\right]dz,\label{L1D}
\end{eqnarray}
where both $\ast$ and c.c. stand for the complex conjugation (as
usual, the derivation neglects terms including $\partial\eta/\partial z$,
assuming that that the variation of transverse width $\eta$ follows
that of\ axial density $|f|^{2}$ \cite{Salasnich_PRA02,Salasnich_PRA02-2}).
Euler-Lagrange equations following from the 1D Lagrangian are 
\begin{equation}
i\frac{\partial f}{\partial t}=-\frac{1}{2}\frac{\partial^{2}f}{\partial z^{2}}+V(z)f+\frac{1}{2\eta^{2}}\left(\frac{1}{4\eta^{2}}-\varepsilon^{3}\right)f+\frac{g|f|^{2}}{4\eta^{4}}f,\label{eq:sig2}
\end{equation}
\begin{equation}
2\varepsilon^{3}\eta^{2}-1-g|f|{}^{2}=0.\label{eq:sig}
\end{equation}
Next, substituting $\eta^{2}$ from Eq. (\ref{eq:sig}) in Eq. (\ref{eq:sig2}),
one arrives at the following time-dependent 1D-NPSE: 
\begin{equation}
i\frac{\partial f}{\partial t}=-\frac{1}{2}\frac{\partial^{2}f}{\partial z^{2}}+V(z)f-\frac{\varepsilon^{6}}{2(1+g|f|^{2})^{2}}f,\label{eq:EFF}
\end{equation}
which is the main result of the derivation. To test accuracy of the
effective 1D-NPSE given by Eq. (\ref{eq:EFF}), below we compare it
to other approximations.

\begin{figure*}[t]
\centering \includegraphics[width=0.8\textwidth]{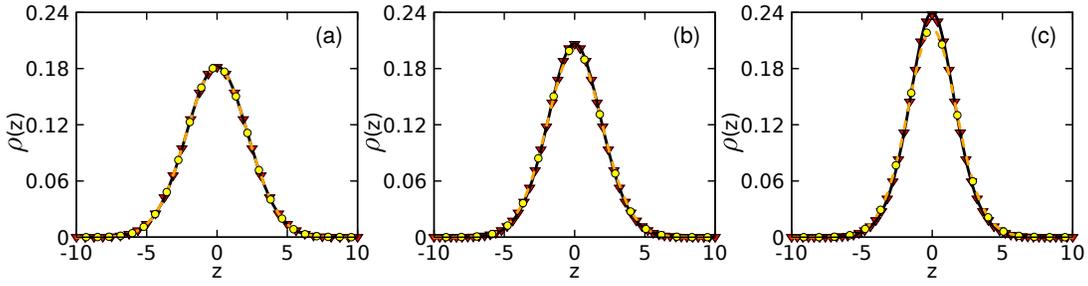} \caption{The same as in Fig. (\ref{F1}), but for the self-attractive BEC $(g<0)$,
for $g=-0.05$ (a), $g=-0.3$ (b), and $g=-0.5$ (c).}
\label{F2} 
\end{figure*}

\subsection{The YXWLH equation}

An effective 1D equation for BEC under the action of the funnel-shaped
potential was proposed in Ref. \cite{Li_CTP09}, in which the dimensional
reduction was performed by means of the VA based on the Gaussian \emph{ansatz}
for the transverse wave function: 
\begin{equation}
\psi(r,z,t)=\exp\left(-\frac{r^{2}}{2\eta^{2}}\right)\frac{f(z,t)}{\sqrt{\pi}\delta},\label{eq:ANZ-1}
\end{equation}
cf. Eq. (\ref{eq:ANZ}). The effective 1D equation derived in Ref.
\cite{Li_CTP09} is 
\begin{equation}
i\frac{\partial f}{\partial t}=-\frac{1}{2}\frac{\partial^{2}f}{\partial z^{2}}+V(z)f-\frac{\pi\varepsilon^{6}}{8(1+g|f|^{2})^{2}}f,\label{YXWLH}
\end{equation}
which we refer to below as the YXWLH equation. Note that its structure
is similar to that of Eq. (\ref{eq:EFF}), the difference amounting
to replacing coefficient $1/2$ in front of the nonlinear terms by
$\pi/8\approx\allowbreak0.393$. However, no solutions of Eq. (\ref{YXWLH})
were reported in Ref. \cite{Li_CTP09}, and the asymptotic structure
of the wave function at $r\rightarrow0$ was not considered either,
cf. Eq. (\ref{rho3}), the latter point being essential in the case
of the singular potential.

\subsection{The weak- and strong-nonlinearity cases}

For the weakly interacting BEC, with $g|f|^{2}\ll1,$ Eq. (\ref{eq:EFF})
reduces to the usual 1D nonlinear Schrödinger equation (NLSE) with
the cubic term, 
\begin{equation}
i\frac{\partial f}{\partial t}=-\frac{1}{2}\frac{\partial^{2}f}{\partial z^{2}}+V(z)f-\varepsilon^{6}\left(\frac{1}{2}+g|f|^{2}\right)f.\label{eq:W}
\end{equation}
On the other hand, the Thomas-Fermi (TF) approximation, which neglects
the spatial derivatives in the GPEs, may be applied in the case of
strong self-repulsion. To this end, we substitute a normalized macroscopic
wave function, $\psi(\mathbf{r},t)=\exp\left(-r/2-i\mu t\right)f/\sqrt{2\pi}$,
in the 3D GPE (note that this expression corresponds to our initial
\emph{ansatz} (\ref{eq:ANZ}) with $\eta=1$) and perform the integration
in the transverse plane, $(x,y)$, neglecting, as usual, the kinetic-energy
terms, the result being 
\begin{equation}
|f(z)|^{2}=\begin{cases}
{\frac{9}{4g}\left[4\mu-4V(z)+\varepsilon^{3}\right],\,\mathrm{at}\,\mu+\varepsilon^{3}>V(z),}\\
{\newline0,\,\mathrm{at}\,\mu+\varepsilon^{3}\leq V(z).}
\end{cases}\label{eq:TF}
\end{equation}
In this approximation, chemical potential $\mu$ is determined by
normalization condition (\ref{N}).

\section{Numerical results \label{NR}}

Numerical results were produced by means of imaginary- and real-time
simulations of the full 3D GPE (\ref{3DGPE_scaled}), as well as 1D
equations (\ref{eq:EFF}), (\ref{YXWLH}), and (\ref{eq:W}). The
simulations were performed with the help of the split-step method,
that was based on the Crank-Nicolson algorithm. The imaginary-time
simulations aimed to produce the system's GS, while the real-time
simulations made it possible to explore dynamical behavior of the
models.

We start the analysis of trapped states, predicted by the 3D and 1D
equations, adopting the HO potential acting in the axial direction,
\begin{equation}
V(z)=\frac{1}{2}\lambda^{2}z^{2},\label{eq:POT1}
\end{equation}
with $\lambda^{2}\ll1$, as the longitudinal potential must be much
weaker than its transverse counterpart.{\LARGE{}{} }To compare the
respective GS solutions produced by the 3D and 1D equations, we use
the axial-density profile of the 3D state, defined as 
\begin{equation}
\rho(z)=\int\int|\psi\left(x,y,z\right)|^{2}dxdy,\label{eq:DENSZ}
\end{equation}
while $\rho(z)=|f(z)|^{2}$ for the 1D equations. In Fig. \ref{F1},
we display typical examples of axial densities for repulsive BEC $(g>0)$
in its GS, under the action of the combined potential (\ref{TPOT}),
in which the longitudinal term is given by Eq. (\ref{eq:POT1}) with
$\lambda=0.1$. These plots demonstrate that the 1D-NPSE provides
essentially more accurate results than other approximations, although
Eq. (\ref{YXWLH}), being close to the 1D-NPSE, leads to accuracy
which is only slightly worse than provided by the main 1D-NPSE approximation.
Thus, it is the best approximation in all cases.

Note that, as observed in Fig. \ref{F1}, the accuracy of the TF approximation
improves slowly with the increase of strength $g$ of the self-repulsion
in the underlying GPE (\ref{3DGPE_scaled}), while it is usually assumed
that the approximation should become better for strong self-repulsion.
There are two reasons for that: first, the TF approximation works
well with positive confining potentials, such as the OH trap (\ref{eq:POT1}),
but not necessarily with the singular negative potential, such as
the funnel one in Eq. (\ref{TPOT}); second, the saturable form of
Eq. (\ref{eq:EFF}) leads to self-cancellation of the strong nonlinearity.

We also analyzed the case of the self-attractive nonlinearity $(g<0)$,
under the action of the same combined potential (\ref{TPOT}). In
Fig. \ref{F2}, we display GS profiles obtained by means of the imaginary-time
propagation, applied to the same set of models, \textit{viz}., the
full 3D GPE, 1D-NPSE, 1D-YWLH equation, and cubic 1D-NLSE, while the
TF approximation is not relevant in the case of the self-attraction.
In this case, all 1D approximations provide good accuracy in comparison
with the full 3D findings.

It is well known that the 3D GPE \cite{Fibich_15} and its 1D reductions
\cite{Salasnich_PRA02,Salasnich_PRA02-2} lead to collapse under
the action of a sufficiently strong self-attraction. By means of systematic
simulations, we have found that the collapse occurs when the attraction
strength exceeds a critical value, $|g|>|g_{c}|$, which is $g_{c}\simeq-0.9$,
$g_{c}\simeq-0.8$, and $g_{c}\simeq-0.9$, for the full 3D GPE, 1D-NPSE,
and 1D-YXWLH equation, respectively, under the action of the funnel
trap (\ref{transverse}) with $\varepsilon=1$ and longitudinal potential
(\ref{eq:POT1}) with $\lambda=0.1$. Although the 1D-NPSE model predicts
$g_{c}$ with a slightly poorer accuracy than the 1D-YXWLH equation
(\ref{YXWLH}), the comparison of the 1D profiles of the wave functions
(not shown here in detail) demonstrates that the former approximation
yields better accuracy at $|g|\leq0.8$.

\begin{figure}[tb]
\centering \includegraphics[width=0.8\columnwidth]{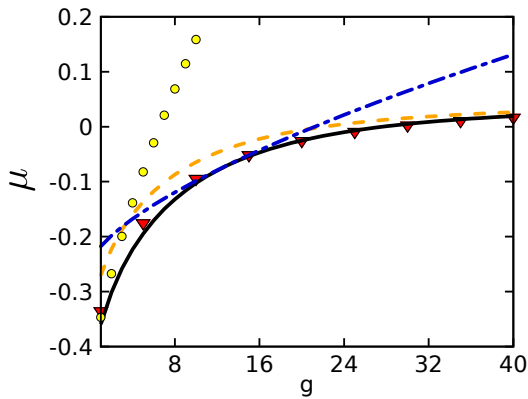} \caption{Numerically found chemical potential $\mu$ of the ground state vs.
$g$ in the self-repulsive BEC $(g>0)$. Parameters and the notation
are the same as in Figs. \ref{F1} and \ref{F2}.}
\label{F3} 
\end{figure}

Getting back to the model with the self-repulsion, in Fig. \ref{F3}
we display the chemical potential, $\mu$, as obtained numerically
from the 3D GPE and from 1D approximations for stationary states,\ which
are looked for, respectively, as $\exp\left(-i\mu t\right)\phi\left(x,y,z\right)$
or $\exp\left(-i\mu t\right)h(z)$, with real functions $\phi$ and
$h$, cf. Eq. (\ref{rho3}). Again, one observes in Fig. \ref{F3}
that the 1D-NPSE provides very accurate results {[}in particular,
conspicuously more accurate than those produced by the similar 1D-YXWLH
approximate equation (\ref{YXWLH}){]}.

\begin{figure}[tb]
\centering \includegraphics[width=0.8\columnwidth]{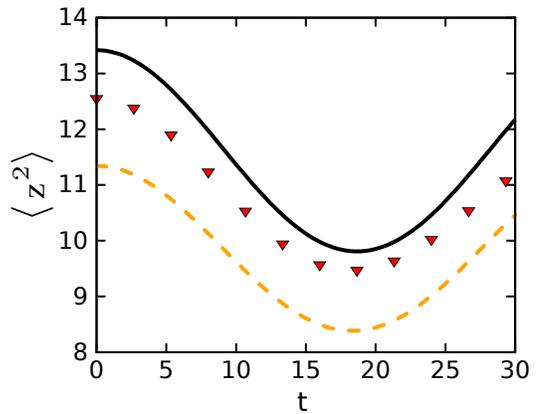} \caption{The evolution of the axial mean-squared size $\left\langle z^{2}\right\rangle $
of the dynamical state generated by the quench, i.e., sudden perturbation
of the ground-state wave function as per Eq. (\ref{1.2}). The self-repulsion
coefficient is $g=5$, other parameters and the notation being the
same as in Figs. \ref{F1}-\ref{F3}. The results produced by the
simplest approximation, in the form of 1D-NLSE (\ref{eq:W}), are
not displayed here, as their difference form the 3D counterparts is
very large.}
\label{F4} 
\end{figure}

Finally, we check the accuracy of the 1D-NPSE in simulations of dynamical
effects. To this end, we used the input in the form of the GS solutions,
which were obtained as outlined above, and ran real-time simulations,
suddenly replacing longitudinal potential (\ref{eq:POT1}) by a slightly
stronger one, 
\begin{equation}
V(z)\rightarrow1.2\times\left(\frac{1}{2}\lambda^{2}z^{2}\right)\label{1.2}
\end{equation}
(the dynamical behavior of this type is often called \textit{quench}
\cite{Dziarmaga_AP10}). This variation of the axial confinement
gives rise to oscillations of the wave function, which can be observed
from the calculation of the axial mean-squared length, $\langle z^{2}\rangle\equiv\int z^{2}|\psi(\mathbf{r})|^{2}d\mathbf{r}$
for 3D GPE, and $\langle z^{2}\rangle\equiv\int_{-\infty}^{+\infty}z^{2}|f|^{2}dz$
for the 1D approximations. In Fig. \ref{F4} we compare the evolution
of $\langle z^{2}\rangle$ as obtained from the simulations of the
full 3D GPE and its 1D counterparts. It is again concluded that the
1D-NPSE is more accurate in comparison with the other approximations.
Note that the simplest cubic model, based on 1D-NLSE, i.e., Eq. (\ref{eq:W}),
is very inaccurate in the application to dynamical effects, therefore
the respective results are not included in Fig. \ref{F4}.

Lastly, in the case of self-repulsion, $g>0$, Eq. (\ref{eq:EFF})
without the axial potential, $V(z)=0$, generates dark solitons. With
this form of the saturable nonlinearity, solutions for dark solitons
where found in an exact but cumbersome analytical form in Ref. \cite{Krolikowski_OL93}.
The comparison of those solutions to their counterparts which may
be produced by the 3D equation (\ref{3DGPE}) is a subject for a separate
work.

\section{Conclusion \label{Conclusion}}

In this work, by applying the dimensional-reduction method, which
is based on the VA (variational approximation), to the full three-dimensional
GPE, we have derived an effective 1D-NPSE (nonpolynomial Schrödinger
equation with rational nonlinearity) that governs the axial mean-field
dynamics of BEC which is tightly trapped in the transverse plane,
with both repulsive and attractive signs of the self-interaction.
We have considered the specific case when the transverse trapping
is imposed by the singular\textit{\ funnel potential}, which is a
singular one, $\sim-1/r$. This potential was originally introduced,
as a model one, in Ref. \cite{Li_CTP09}. We propose physical settings
which give rise to this model, \textit{viz}., a uniformly charged
axial wire attracting particles carrying a permanent dipole electric
moment, as well as a bosonic gas of magnetic atoms pulled to the axial
electric current. In spite of the singularity, the funnel potential
maintains regular wave functions of the trapped states. We verify
the accuracy of predictions provided by the 1D-NPSE by comparing them
to results of the full 3D simulations, and other 1D approximations,
such as the usual cubic one-dimensional NLSE and the Thomas--Fermi
approximation. In the case of self-attraction ($g<0$), the setting
gives rise to collapse at $|g|>\left\vert g_{c}\right\vert $, with
the critical value also accurately predicted by the 1D reduction.
Finally, the 1D-NPSE also provides a sufficiently accurate approximation
for dynamical regimes, initiated by a sudden quench applied to the
axial potential. Thus, the approximation based on the 1D equation
of the NPSE type is relevant for various types of the transverse trapping
potentials, including singular ones, and for the analysis of both
the systems' ground states and dynamical regimes.

Finally, it may be interesting to generalize the above analysis for
modes with embedded vorticity $S=1,2,3,...$, represented by factor
$\exp\left(iS\theta\right)$ in the 3D solution, where $\theta$ is
the angular variable in the transverse plane \cite{Salasnich_PRA07.063614}.
Then, the expansion similar to that in Eq. (\ref{rho3}) starts with
terms $\psi=\exp\left(-i\mu t+iS\theta\right)r^{S}\left\{ \left[1-\varepsilon^{3}\left(1+2S\right)^{-1}r\right]\psi_{0}(z)+...\right\} $.
This extension of the analysis will be reported elsewhere.

\section*{Acknowledgments}

We appreciate valuable discussions with Luca Salasnich. Financial
support from the Brazilian agencies CNPq (\#304073/2016-4 \& \#425718/2018-2),
CAPES, and FAPEG (PRONEM \#201710267000540 \& PRONEX \#201710267000503)
is acknowledged. This work was performed as part of the Brazilian
National Institute of Science and Technology (INCT) for Quantum Information
(\#465469/2014-0). The work B.A.M. is supported, in part, by the joint
program in physics between NSF and Binational (US-Israel) Science
Foundation through project No. 2015616, by the Israel Science Foundation
through Grant No. 1286/17, and by CAPES\
(Brazil) through program PRINT, grant No. 88887.364746/2019-00.


\begin{thebibliography}{68}%
\makeatletter
\providecommand \@ifxundefined [1]{%
 \@ifx{#1\undefined}
}%
\providecommand \@ifnum [1]{%
 \ifnum #1\expandafter \@firstoftwo
 \else \expandafter \@secondoftwo
 \fi
}%
\providecommand \@ifx [1]{%
 \ifx #1\expandafter \@firstoftwo
 \else \expandafter \@secondoftwo
 \fi
}%
\providecommand \natexlab [1]{#1}%
\providecommand \enquote  [1]{``#1''}%
\providecommand \bibnamefont  [1]{#1}%
\providecommand \bibfnamefont [1]{#1}%
\providecommand \citenamefont [1]{#1}%
\providecommand \href@noop [0]{\@secondoftwo}%
\providecommand \href [0]{\begingroup \@sanitize@url \@href}%
\providecommand \@href[1]{\@@startlink{#1}\@@href}%
\providecommand \@@href[1]{\endgroup#1\@@endlink}%
\providecommand \@sanitize@url [0]{\catcode `\\12\catcode `\$12\catcode
  `\&12\catcode `\#12\catcode `\^12\catcode `\_12\catcode `\%12\relax}%
\providecommand \@@startlink[1]{}%
\providecommand \@@endlink[0]{}%
\providecommand \url  [0]{\begingroup\@sanitize@url \@url }%
\providecommand \@url [1]{\endgroup\@href {#1}{\urlprefix }}%
\providecommand \urlprefix  [0]{URL }%
\providecommand \Eprint [0]{\href }%
\providecommand \doibase [0]{http://dx.doi.org/}%
\providecommand \selectlanguage [0]{\@gobble}%
\providecommand \bibinfo  [0]{\@secondoftwo}%
\providecommand \bibfield  [0]{\@secondoftwo}%
\providecommand \translation [1]{[#1]}%
\providecommand \BibitemOpen [0]{}%
\providecommand \bibitemStop [0]{}%
\providecommand \bibitemNoStop [0]{.\EOS\space}%
\providecommand \EOS [0]{\spacefactor3000\relax}%
\providecommand \BibitemShut  [1]{\csname bibitem#1\endcsname}%
\let\auto@bib@innerbib\@empty
\bibitem [{\citenamefont {Anderson}\ \emph {et~al.}(1995)\citenamefont
  {Anderson}, \citenamefont {Ensher}, \citenamefont {Matthews}, \citenamefont
  {Wieman},\ and\ \citenamefont {Cornell}}]{Anderson_SCI95}%
  \BibitemOpen
  \bibfield  {author} {\bibinfo {author} {\bibfnamefont {M.~H.}\ \bibnamefont
  {Anderson}}, \bibinfo {author} {\bibfnamefont {J.~R.}\ \bibnamefont
  {Ensher}}, \bibinfo {author} {\bibfnamefont {M.~R.}\ \bibnamefont
  {Matthews}}, \bibinfo {author} {\bibfnamefont {C.~E.}\ \bibnamefont
  {Wieman}}, \ and\ \bibinfo {author} {\bibfnamefont {E.~A.}\ \bibnamefont
  {Cornell}},\ }\href {\doibase 10.1126/science.269.5221.198} {\bibfield
  {journal} {\bibinfo  {journal} {Science}\ }\textbf {\bibinfo {volume}
  {269}},\ \bibinfo {pages} {198} (\bibinfo {year} {1995})}\BibitemShut
  {NoStop}%
\bibitem [{\citenamefont {Davis}\ \emph {et~al.}(1995)\citenamefont {Davis},
  \citenamefont {Mewes}, \citenamefont {Andrews}, \citenamefont {van Druten},
  \citenamefont {Durfee}, \citenamefont {Kurn},\ and\ \citenamefont
  {Ketterle}}]{Davis_PRL95}%
  \BibitemOpen
  \bibfield  {author} {\bibinfo {author} {\bibfnamefont {K.~B.}\ \bibnamefont
  {Davis}}, \bibinfo {author} {\bibfnamefont {M.~O.}\ \bibnamefont {Mewes}},
  \bibinfo {author} {\bibfnamefont {M.~R.}\ \bibnamefont {Andrews}}, \bibinfo
  {author} {\bibfnamefont {N.~J.}\ \bibnamefont {van Druten}}, \bibinfo
  {author} {\bibfnamefont {D.~S.}\ \bibnamefont {Durfee}}, \bibinfo {author}
  {\bibfnamefont {D.~M.}\ \bibnamefont {Kurn}}, \ and\ \bibinfo {author}
  {\bibfnamefont {W.}~\bibnamefont {Ketterle}},\ }\href {\doibase
  10.1103/PhysRevLett.75.3969} {\bibfield  {journal} {\bibinfo  {journal}
  {Physical Review Letters}\ }\textbf {\bibinfo {volume} {75}},\ \bibinfo
  {pages} {3969} (\bibinfo {year} {1995})}\BibitemShut {NoStop}%
\bibitem [{\citenamefont {Bradley}\ \emph {et~al.}(1997)\citenamefont
  {Bradley}, \citenamefont {Sackett},\ and\ \citenamefont
  {Hulet}}]{Bradley.78.985}%
  \BibitemOpen
  \bibfield  {author} {\bibinfo {author} {\bibfnamefont {C.~C.}\ \bibnamefont
  {Bradley}}, \bibinfo {author} {\bibfnamefont {C.~A.}\ \bibnamefont
  {Sackett}}, \ and\ \bibinfo {author} {\bibfnamefont {R.~G.}\ \bibnamefont
  {Hulet}},\ }\href {\doibase 10.1103/PhysRevLett.78.985} {\bibfield  {journal}
  {\bibinfo  {journal} {Phys. Rev. Lett.}\ }\textbf {\bibinfo {volume} {78}},\
  \bibinfo {pages} {985} (\bibinfo {year} {1997})}\BibitemShut {NoStop}%
\bibitem [{\citenamefont {Griesmaier}\ \emph {et~al.}(2005)\citenamefont
  {Griesmaier}, \citenamefont {Werner}, \citenamefont {Hensler}, \citenamefont
  {Stuhler},\ and\ \citenamefont {Pfau}}]{Gries.94.160401}%
  \BibitemOpen
  \bibfield  {author} {\bibinfo {author} {\bibfnamefont {A.}~\bibnamefont
  {Griesmaier}}, \bibinfo {author} {\bibfnamefont {J.}~\bibnamefont {Werner}},
  \bibinfo {author} {\bibfnamefont {S.}~\bibnamefont {Hensler}}, \bibinfo
  {author} {\bibfnamefont {J.}~\bibnamefont {Stuhler}}, \ and\ \bibinfo
  {author} {\bibfnamefont {T.}~\bibnamefont {Pfau}},\ }\href {\doibase
  10.1103/PhysRevLett.94.160401} {\bibfield  {journal} {\bibinfo  {journal}
  {Phys. Rev. Lett.}\ }\textbf {\bibinfo {volume} {94}},\ \bibinfo {pages}
  {160401} (\bibinfo {year} {2005})}\BibitemShut {NoStop}%
\bibitem [{\citenamefont {Fried}\ \emph {et~al.}(1998)\citenamefont {Fried},
  \citenamefont {Killian}, \citenamefont {Willmann}, \citenamefont {Landhuis},
  \citenamefont {Moss}, \citenamefont {Kleppner},\ and\ \citenamefont
  {Greytak}}]{Fried_PRL98}%
  \BibitemOpen
  \bibfield  {author} {\bibinfo {author} {\bibfnamefont {D.~G.}\ \bibnamefont
  {Fried}}, \bibinfo {author} {\bibfnamefont {T.~C.}\ \bibnamefont {Killian}},
  \bibinfo {author} {\bibfnamefont {L.}~\bibnamefont {Willmann}}, \bibinfo
  {author} {\bibfnamefont {D.}~\bibnamefont {Landhuis}}, \bibinfo {author}
  {\bibfnamefont {S.~C.}\ \bibnamefont {Moss}}, \bibinfo {author}
  {\bibfnamefont {D.}~\bibnamefont {Kleppner}}, \ and\ \bibinfo {author}
  {\bibfnamefont {T.~J.}\ \bibnamefont {Greytak}},\ }\href {\doibase
  10.1103/PhysRevLett.81.3811} {\bibfield  {journal} {\bibinfo  {journal}
  {Physical Review Letters}\ }\textbf {\bibinfo {volume} {81}},\ \bibinfo
  {pages} {3811} (\bibinfo {year} {1998})}\BibitemShut {NoStop}%
\bibitem [{\citenamefont {Aikawa}\ \emph {et~al.}(2012)\citenamefont {Aikawa},
  \citenamefont {Frisch}, \citenamefont {Mark}, \citenamefont {Baier},
  \citenamefont {Rietzler}, \citenamefont {Grimm},\ and\ \citenamefont
  {Ferlaino}}]{Aika.108.210401}%
  \BibitemOpen
  \bibfield  {author} {\bibinfo {author} {\bibfnamefont {K.}~\bibnamefont
  {Aikawa}}, \bibinfo {author} {\bibfnamefont {A.}~\bibnamefont {Frisch}},
  \bibinfo {author} {\bibfnamefont {M.}~\bibnamefont {Mark}}, \bibinfo {author}
  {\bibfnamefont {S.}~\bibnamefont {Baier}}, \bibinfo {author} {\bibfnamefont
  {A.}~\bibnamefont {Rietzler}}, \bibinfo {author} {\bibfnamefont
  {R.}~\bibnamefont {Grimm}}, \ and\ \bibinfo {author} {\bibfnamefont
  {F.}~\bibnamefont {Ferlaino}},\ }\href {\doibase
  10.1103/PhysRevLett.108.210401} {\bibfield  {journal} {\bibinfo  {journal}
  {Phys. Rev. Lett.}\ }\textbf {\bibinfo {volume} {108}},\ \bibinfo {pages}
  {210401} (\bibinfo {year} {2012})}\BibitemShut {NoStop}%
\bibitem [{\citenamefont {Jochim}\ \emph {et~al.}(2003)\citenamefont {Jochim},
  \citenamefont {Bartenstein}, \citenamefont {Altmeyer}, \citenamefont {Hendl},
  \citenamefont {Riedl}, \citenamefont {Chin}, \citenamefont {Denschlag},\ and\
  \citenamefont {Grimm}}]{Jochim_SCI03}%
  \BibitemOpen
  \bibfield  {author} {\bibinfo {author} {\bibfnamefont {S.}~\bibnamefont
  {Jochim}}, \bibinfo {author} {\bibfnamefont {M.}~\bibnamefont {Bartenstein}},
  \bibinfo {author} {\bibfnamefont {A.}~\bibnamefont {Altmeyer}}, \bibinfo
  {author} {\bibfnamefont {G.}~\bibnamefont {Hendl}}, \bibinfo {author}
  {\bibfnamefont {S.}~\bibnamefont {Riedl}}, \bibinfo {author} {\bibfnamefont
  {C.}~\bibnamefont {Chin}}, \bibinfo {author} {\bibfnamefont {J.~H.}\
  \bibnamefont {Denschlag}}, \ and\ \bibinfo {author} {\bibfnamefont
  {R.}~\bibnamefont {Grimm}},\ }\href {\doibase 10.1126/science.1093280}
  {\bibfield  {journal} {\bibinfo  {journal} {Science}\ }\textbf {\bibinfo
  {volume} {302}},\ \bibinfo {pages} {2101} (\bibinfo {year}
  {2003})}\BibitemShut {NoStop}%
\bibitem [{\citenamefont {Zwierlein}\ \emph {et~al.}(2003)\citenamefont
  {Zwierlein}, \citenamefont {Stan}, \citenamefont {Schunck}, \citenamefont
  {Raupach}, \citenamefont {Gupta}, \citenamefont {Hadzibabic},\ and\
  \citenamefont {Ketterle}}]{Zwierlein.91.250401}%
  \BibitemOpen
  \bibfield  {author} {\bibinfo {author} {\bibfnamefont {M.~W.}\ \bibnamefont
  {Zwierlein}}, \bibinfo {author} {\bibfnamefont {C.~A.}\ \bibnamefont {Stan}},
  \bibinfo {author} {\bibfnamefont {C.~H.}\ \bibnamefont {Schunck}}, \bibinfo
  {author} {\bibfnamefont {S.~M.~F.}\ \bibnamefont {Raupach}}, \bibinfo
  {author} {\bibfnamefont {S.}~\bibnamefont {Gupta}}, \bibinfo {author}
  {\bibfnamefont {Z.}~\bibnamefont {Hadzibabic}}, \ and\ \bibinfo {author}
  {\bibfnamefont {W.}~\bibnamefont {Ketterle}},\ }\href {\doibase
  10.1103/PhysRevLett.91.250401} {\bibfield  {journal} {\bibinfo  {journal}
  {Phys. Rev. Lett.}\ }\textbf {\bibinfo {volume} {91}},\ \bibinfo {pages}
  {250401} (\bibinfo {year} {2003})}\BibitemShut {NoStop}%
\bibitem [{\citenamefont {Burger}\ \emph {et~al.}(1999)\citenamefont {Burger},
  \citenamefont {Bongs}, \citenamefont {Dettmer}, \citenamefont {Ertmer},
  \citenamefont {Sengstock}, \citenamefont {Sanpera}, \citenamefont
  {Shlyapnikov},\ and\ \citenamefont {Lewenstein}}]{Burger_PRL99}%
  \BibitemOpen
  \bibfield  {author} {\bibinfo {author} {\bibfnamefont {S.}~\bibnamefont
  {Burger}}, \bibinfo {author} {\bibfnamefont {K.}~\bibnamefont {Bongs}},
  \bibinfo {author} {\bibfnamefont {S.}~\bibnamefont {Dettmer}}, \bibinfo
  {author} {\bibfnamefont {W.}~\bibnamefont {Ertmer}}, \bibinfo {author}
  {\bibfnamefont {K.}~\bibnamefont {Sengstock}}, \bibinfo {author}
  {\bibfnamefont {A.}~\bibnamefont {Sanpera}}, \bibinfo {author} {\bibfnamefont
  {G.~V.}\ \bibnamefont {Shlyapnikov}}, \ and\ \bibinfo {author} {\bibfnamefont
  {M.}~\bibnamefont {Lewenstein}},\ }\href {\doibase
  10.1103/PhysRevLett.83.5198} {\bibfield  {journal} {\bibinfo  {journal}
  {Physical Review Letters}\ }\textbf {\bibinfo {volume} {83}},\ \bibinfo
  {pages} {5198} (\bibinfo {year} {1999})}\BibitemShut {NoStop}%
\bibitem [{\citenamefont {Denschlag}\ \emph {et~al.}(2000)\citenamefont
  {Denschlag}, \citenamefont {Simsarian}, \citenamefont {Feder}, \citenamefont
  {Clark}, \citenamefont {Collins}, \citenamefont {Cubizolles}, \citenamefont
  {Deng}, \citenamefont {Hagley}, \citenamefont {Helmerson}, \citenamefont
  {Reinhardt}, \citenamefont {Rolston}, \citenamefont {Schneider},\ and\
  \citenamefont {Phillips}}]{Denschlag_SCI00}%
  \BibitemOpen
  \bibfield  {author} {\bibinfo {author} {\bibfnamefont {J.}~\bibnamefont
  {Denschlag}}, \bibinfo {author} {\bibfnamefont {J.~E.}\ \bibnamefont
  {Simsarian}}, \bibinfo {author} {\bibfnamefont {D.~L.}\ \bibnamefont
  {Feder}}, \bibinfo {author} {\bibfnamefont {C.~W.}\ \bibnamefont {Clark}},
  \bibinfo {author} {\bibfnamefont {L.~A.}\ \bibnamefont {Collins}}, \bibinfo
  {author} {\bibfnamefont {J.}~\bibnamefont {Cubizolles}}, \bibinfo {author}
  {\bibfnamefont {L.}~\bibnamefont {Deng}}, \bibinfo {author} {\bibfnamefont
  {E.~W.}\ \bibnamefont {Hagley}}, \bibinfo {author} {\bibfnamefont
  {K.}~\bibnamefont {Helmerson}}, \bibinfo {author} {\bibfnamefont {W.~P.}\
  \bibnamefont {Reinhardt}}, \bibinfo {author} {\bibfnamefont {S.~L.}\
  \bibnamefont {Rolston}}, \bibinfo {author} {\bibfnamefont {B.~I.}\
  \bibnamefont {Schneider}}, \ and\ \bibinfo {author} {\bibfnamefont {W.~D.}\
  \bibnamefont {Phillips}},\ }\href {\doibase 10.1126/science.287.5450.97}
  {\bibfield  {journal} {\bibinfo  {journal} {Science}\ }\textbf {\bibinfo
  {volume} {287}},\ \bibinfo {pages} {97} (\bibinfo {year} {2000})}\BibitemShut
  {NoStop}%
\bibitem [{\citenamefont {Khaykovich}\ \emph {et~al.}(2002)\citenamefont
  {Khaykovich}, \citenamefont {Schreck}, \citenamefont {Ferrari}, \citenamefont
  {Bourdel}, \citenamefont {Cubizolles}, \citenamefont {Carr}, \citenamefont
  {Castin},\ and\ \citenamefont {Salomon}}]{Khaykovich_SCI02}%
  \BibitemOpen
  \bibfield  {author} {\bibinfo {author} {\bibfnamefont {L.}~\bibnamefont
  {Khaykovich}}, \bibinfo {author} {\bibfnamefont {F.}~\bibnamefont {Schreck}},
  \bibinfo {author} {\bibfnamefont {G.}~\bibnamefont {Ferrari}}, \bibinfo
  {author} {\bibfnamefont {T.}~\bibnamefont {Bourdel}}, \bibinfo {author}
  {\bibfnamefont {J.}~\bibnamefont {Cubizolles}}, \bibinfo {author}
  {\bibfnamefont {L.~D.}\ \bibnamefont {Carr}}, \bibinfo {author}
  {\bibfnamefont {Y.}~\bibnamefont {Castin}}, \ and\ \bibinfo {author}
  {\bibfnamefont {C.}~\bibnamefont {Salomon}},\ }\href {\doibase
  10.1126/science.1071021} {\bibfield  {journal} {\bibinfo  {journal}
  {Science}\ }\textbf {\bibinfo {volume} {296}},\ \bibinfo {pages} {1290}
  (\bibinfo {year} {2002})}\BibitemShut {NoStop}%
\bibitem [{\citenamefont {Strecker}\ \emph {et~al.}(2002)\citenamefont
  {Strecker}, \citenamefont {Partridge}, \citenamefont {Truscott},\ and\
  \citenamefont {Hulet}}]{Strecker_NAT02}%
  \BibitemOpen
  \bibfield  {author} {\bibinfo {author} {\bibfnamefont {K.~E.}\ \bibnamefont
  {Strecker}}, \bibinfo {author} {\bibfnamefont {G.~B.}\ \bibnamefont
  {Partridge}}, \bibinfo {author} {\bibfnamefont {A.~G.}\ \bibnamefont
  {Truscott}}, \ and\ \bibinfo {author} {\bibfnamefont {R.~G.}\ \bibnamefont
  {Hulet}},\ }\href {\doibase 10.1038/nature747} {\bibfield  {journal}
  {\bibinfo  {journal} {Nature}\ }\textbf {\bibinfo {volume} {417}},\ \bibinfo
  {pages} {150} (\bibinfo {year} {2002})}\BibitemShut {NoStop}%
\bibitem [{\citenamefont {Matthews}\ \emph {et~al.}(1999)\citenamefont
  {Matthews}, \citenamefont {Anderson}, \citenamefont {Haljan}, \citenamefont
  {Hall}, \citenamefont {Wieman},\ and\ \citenamefont
  {Cornell}}]{Matthews_PRL99}%
  \BibitemOpen
  \bibfield  {author} {\bibinfo {author} {\bibfnamefont {M.~R.}\ \bibnamefont
  {Matthews}}, \bibinfo {author} {\bibfnamefont {B.~P.}\ \bibnamefont
  {Anderson}}, \bibinfo {author} {\bibfnamefont {P.~C.}\ \bibnamefont
  {Haljan}}, \bibinfo {author} {\bibfnamefont {D.~S.}\ \bibnamefont {Hall}},
  \bibinfo {author} {\bibfnamefont {C.~E.}\ \bibnamefont {Wieman}}, \ and\
  \bibinfo {author} {\bibfnamefont {E.~A.}\ \bibnamefont {Cornell}},\ }\href
  {\doibase 10.1103/PhysRevLett.83.2498} {\bibfield  {journal} {\bibinfo
  {journal} {Physical Review Letters}\ }\textbf {\bibinfo {volume} {83}},\
  \bibinfo {pages} {2498} (\bibinfo {year} {1999})}\BibitemShut {NoStop}%
\bibitem [{\citenamefont {Madison}\ \emph {et~al.}(2000)\citenamefont
  {Madison}, \citenamefont {Chevy}, \citenamefont {Wohlleben},\ and\
  \citenamefont {Dalibard}}]{Madison_PRL00}%
  \BibitemOpen
  \bibfield  {author} {\bibinfo {author} {\bibfnamefont {K.~W.}\ \bibnamefont
  {Madison}}, \bibinfo {author} {\bibfnamefont {F.}~\bibnamefont {Chevy}},
  \bibinfo {author} {\bibfnamefont {W.}~\bibnamefont {Wohlleben}}, \ and\
  \bibinfo {author} {\bibfnamefont {J.}~\bibnamefont {Dalibard}},\ }\href
  {\doibase 10.1103/PhysRevLett.84.806} {\bibfield  {journal} {\bibinfo
  {journal} {Physical Review Letters}\ }\textbf {\bibinfo {volume} {84}},\
  \bibinfo {pages} {806} (\bibinfo {year} {2000})}\BibitemShut {NoStop}%
\bibitem [{\citenamefont {Malomed}(2019)}]{Malomed_arxiv19}%
  \BibitemOpen
  \bibfield  {author} {\bibinfo {author} {\bibfnamefont {B.~A.}\ \bibnamefont
  {Malomed}},\ }\href {http://arxiv.org/abs/1904.12081} {\  (\bibinfo {year}
  {2019})},\ \Eprint {http://arxiv.org/abs/1904.12081} {arXiv:1904.12081}
  \BibitemShut {NoStop}%
\bibitem [{\citenamefont {Inouye}\ \emph {et~al.}(1998)\citenamefont {Inouye},
  \citenamefont {Andrews}, \citenamefont {Stenger}, \citenamefont {Miesner},
  \citenamefont {Stamper-Kurn},\ and\ \citenamefont {Ketterle}}]{Inouye_NAT98}%
  \BibitemOpen
  \bibfield  {author} {\bibinfo {author} {\bibfnamefont {S.}~\bibnamefont
  {Inouye}}, \bibinfo {author} {\bibfnamefont {M.~R.}\ \bibnamefont {Andrews}},
  \bibinfo {author} {\bibfnamefont {J.}~\bibnamefont {Stenger}}, \bibinfo
  {author} {\bibfnamefont {H.-J.}\ \bibnamefont {Miesner}}, \bibinfo {author}
  {\bibfnamefont {D.~M.}\ \bibnamefont {Stamper-Kurn}}, \ and\ \bibinfo
  {author} {\bibfnamefont {W.}~\bibnamefont {Ketterle}},\ }\href {\doibase
  10.1038/32354} {\bibfield  {journal} {\bibinfo  {journal} {Nature}\ }\textbf
  {\bibinfo {volume} {392}},\ \bibinfo {pages} {151} (\bibinfo {year}
  {1998})}\BibitemShut {NoStop}%
\bibitem [{\citenamefont {Petrov}(2015)}]{Petrov_PRL15}%
  \BibitemOpen
  \bibfield  {author} {\bibinfo {author} {\bibfnamefont {D.~S.}\ \bibnamefont
  {Petrov}},\ }\href {\doibase 10.1103/PhysRevLett.115.155302} {\bibfield
  {journal} {\bibinfo  {journal} {Physical Review Letters}\ }\textbf {\bibinfo
  {volume} {115}},\ \bibinfo {pages} {155302} (\bibinfo {year}
  {2015})}\BibitemShut {NoStop}%
\bibitem [{\citenamefont {Petrov}\ and\ \citenamefont
  {Astrakharchik}(2016)}]{Petrov_PRL16}%
  \BibitemOpen
  \bibfield  {author} {\bibinfo {author} {\bibfnamefont {D.~S.}\ \bibnamefont
  {Petrov}}\ and\ \bibinfo {author} {\bibfnamefont {G.~E.}\ \bibnamefont
  {Astrakharchik}},\ }\href {\doibase 10.1103/PhysRevLett.117.100401}
  {\bibfield  {journal} {\bibinfo  {journal} {Physical Review Letters}\
  }\textbf {\bibinfo {volume} {117}},\ \bibinfo {pages} {100401} (\bibinfo
  {year} {2016})}\BibitemShut {NoStop}%
\bibitem [{\citenamefont {Cabrera}\ \emph {et~al.}(2018)\citenamefont
  {Cabrera}, \citenamefont {Tanzi}, \citenamefont {Sanz}, \citenamefont
  {Naylor}, \citenamefont {Thomas}, \citenamefont {Cheiney},\ and\
  \citenamefont {Tarruell}}]{Cabrera_SCI18}%
  \BibitemOpen
  \bibfield  {author} {\bibinfo {author} {\bibfnamefont {C.~R.}\ \bibnamefont
  {Cabrera}}, \bibinfo {author} {\bibfnamefont {L.}~\bibnamefont {Tanzi}},
  \bibinfo {author} {\bibfnamefont {J.}~\bibnamefont {Sanz}}, \bibinfo {author}
  {\bibfnamefont {B.}~\bibnamefont {Naylor}}, \bibinfo {author} {\bibfnamefont
  {P.}~\bibnamefont {Thomas}}, \bibinfo {author} {\bibfnamefont
  {P.}~\bibnamefont {Cheiney}}, \ and\ \bibinfo {author} {\bibfnamefont
  {L.}~\bibnamefont {Tarruell}},\ }\href {\doibase 10.1126/science.aao5686}
  {\bibfield  {journal} {\bibinfo  {journal} {Science}\ }\textbf {\bibinfo
  {volume} {359}},\ \bibinfo {pages} {301} (\bibinfo {year}
  {2018})}\BibitemShut {NoStop}%
\bibitem [{\citenamefont {Cheiney}\ \emph {et~al.}(2018)\citenamefont
  {Cheiney}, \citenamefont {Cabrera}, \citenamefont {Sanz}, \citenamefont
  {Naylor}, \citenamefont {Tanzi},\ and\ \citenamefont
  {Tarruell}}]{Cheiney_PRL18}%
  \BibitemOpen
  \bibfield  {author} {\bibinfo {author} {\bibfnamefont {P.}~\bibnamefont
  {Cheiney}}, \bibinfo {author} {\bibfnamefont {C.~R.}\ \bibnamefont
  {Cabrera}}, \bibinfo {author} {\bibfnamefont {J.}~\bibnamefont {Sanz}},
  \bibinfo {author} {\bibfnamefont {B.}~\bibnamefont {Naylor}}, \bibinfo
  {author} {\bibfnamefont {L.}~\bibnamefont {Tanzi}}, \ and\ \bibinfo {author}
  {\bibfnamefont {L.}~\bibnamefont {Tarruell}},\ }\href {\doibase
  10.1103/PhysRevLett.120.135301} {\bibfield  {journal} {\bibinfo  {journal}
  {Physical Review Letters}\ }\textbf {\bibinfo {volume} {120}},\ \bibinfo
  {pages} {135301} (\bibinfo {year} {2018})}\BibitemShut {NoStop}%
\bibitem [{\citenamefont {Semeghini}\ \emph {et~al.}(2018)\citenamefont
  {Semeghini}, \citenamefont {Ferioli}, \citenamefont {Masi}, \citenamefont
  {Mazzinghi}, \citenamefont {Wolswijk}, \citenamefont {Minardi}, \citenamefont
  {Modugno}, \citenamefont {Modugno}, \citenamefont {Inguscio},\ and\
  \citenamefont {Fattori}}]{Semeghini_PRL18}%
  \BibitemOpen
  \bibfield  {author} {\bibinfo {author} {\bibfnamefont {G.}~\bibnamefont
  {Semeghini}}, \bibinfo {author} {\bibfnamefont {G.}~\bibnamefont {Ferioli}},
  \bibinfo {author} {\bibfnamefont {L.}~\bibnamefont {Masi}}, \bibinfo {author}
  {\bibfnamefont {C.}~\bibnamefont {Mazzinghi}}, \bibinfo {author}
  {\bibfnamefont {L.}~\bibnamefont {Wolswijk}}, \bibinfo {author}
  {\bibfnamefont {F.}~\bibnamefont {Minardi}}, \bibinfo {author} {\bibfnamefont
  {M.}~\bibnamefont {Modugno}}, \bibinfo {author} {\bibfnamefont
  {G.}~\bibnamefont {Modugno}}, \bibinfo {author} {\bibfnamefont
  {M.}~\bibnamefont {Inguscio}}, \ and\ \bibinfo {author} {\bibfnamefont
  {M.}~\bibnamefont {Fattori}},\ }\href {\doibase
  10.1103/PhysRevLett.120.235301} {\bibfield  {journal} {\bibinfo  {journal}
  {Physical Review Letters}\ }\textbf {\bibinfo {volume} {120}},\ \bibinfo
  {pages} {235301} (\bibinfo {year} {2018})}\BibitemShut {NoStop}%
\bibitem [{\citenamefont {Ferioli}\ \emph {et~al.}(2019)\citenamefont
  {Ferioli}, \citenamefont {Semeghini}, \citenamefont {Masi}, \citenamefont
  {Giusti}, \citenamefont {Modugno}, \citenamefont {Inguscio}, \citenamefont
  {Gallem{\'{i}}}, \citenamefont {Recati},\ and\ \citenamefont
  {Fattori}}]{Ferioli_PRL19}%
  \BibitemOpen
  \bibfield  {author} {\bibinfo {author} {\bibfnamefont {G.}~\bibnamefont
  {Ferioli}}, \bibinfo {author} {\bibfnamefont {G.}~\bibnamefont {Semeghini}},
  \bibinfo {author} {\bibfnamefont {L.}~\bibnamefont {Masi}}, \bibinfo {author}
  {\bibfnamefont {G.}~\bibnamefont {Giusti}}, \bibinfo {author} {\bibfnamefont
  {G.}~\bibnamefont {Modugno}}, \bibinfo {author} {\bibfnamefont
  {M.}~\bibnamefont {Inguscio}}, \bibinfo {author} {\bibfnamefont
  {A.}~\bibnamefont {Gallem{\'{i}}}}, \bibinfo {author} {\bibfnamefont
  {A.}~\bibnamefont {Recati}}, \ and\ \bibinfo {author} {\bibfnamefont
  {M.}~\bibnamefont {Fattori}},\ }\href {\doibase
  10.1103/PhysRevLett.122.090401} {\bibfield  {journal} {\bibinfo  {journal}
  {Physical Review Letters}\ }\textbf {\bibinfo {volume} {122}},\ \bibinfo
  {pages} {090401} (\bibinfo {year} {2019})}\BibitemShut {NoStop}%
\bibitem [{\citenamefont {D'Errico}\ \emph {et~al.}(2019)\citenamefont
  {D'Errico}, \citenamefont {Burchianti}, \citenamefont {Prevedelli},
  \citenamefont {Salasnich}, \citenamefont {Ancilotto}, \citenamefont
  {Modugno}, \citenamefont {Minardi},\ and\ \citenamefont
  {Fort}}]{D'Errico_ArXiv19}%
  \BibitemOpen
  \bibfield  {author} {\bibinfo {author} {\bibfnamefont {C.}~\bibnamefont
  {D'Errico}}, \bibinfo {author} {\bibfnamefont {A.}~\bibnamefont
  {Burchianti}}, \bibinfo {author} {\bibfnamefont {M.}~\bibnamefont
  {Prevedelli}}, \bibinfo {author} {\bibfnamefont {L.}~\bibnamefont
  {Salasnich}}, \bibinfo {author} {\bibfnamefont {F.}~\bibnamefont
  {Ancilotto}}, \bibinfo {author} {\bibfnamefont {M.}~\bibnamefont {Modugno}},
  \bibinfo {author} {\bibfnamefont {F.}~\bibnamefont {Minardi}}, \ and\
  \bibinfo {author} {\bibfnamefont {C.}~\bibnamefont {Fort}},\ }\href
  {http://arxiv.org/abs/1908.00761} {\  (\bibinfo {year} {2019})},\ \Eprint
  {http://arxiv.org/abs/1908.00761} {arXiv:1908.00761} \BibitemShut {NoStop}%
\bibitem [{\citenamefont {Pitaevskii}\ and\ \citenamefont
  {Stringari}(2003)}]{Pitaevskii_03}%
  \BibitemOpen
  \bibfield  {author} {\bibinfo {author} {\bibfnamefont {L.~P.}\ \bibnamefont
  {Pitaevskii}}\ and\ \bibinfo {author} {\bibfnamefont {S.}~\bibnamefont
  {Stringari}},\ }\href {https://books.google.com.br/books?id=rIobbOxC4j4C}
  {\emph {\bibinfo {title} {{Bose-Einstein Condensation}}}},\ International
  Series of Monographs on Physics\ (\bibinfo  {publisher} {Clarendon Press},\
  \bibinfo {year} {2003})\BibitemShut {NoStop}%
\bibitem [{\citenamefont {Pethick}\ and\ \citenamefont
  {Smith}(2008)}]{Pethick_08}%
  \BibitemOpen
  \bibfield  {author} {\bibinfo {author} {\bibfnamefont {C.~J.}\ \bibnamefont
  {Pethick}}\ and\ \bibinfo {author} {\bibfnamefont {H.}~\bibnamefont
  {Smith}},\ }\href {\doibase 10.1017/CBO9780511802850} {\emph {\bibinfo
  {title} {{Bose-Einstein Condensation in Dilute Gases}}}}\ (\bibinfo
  {publisher} {Cambridge University Press},\ \bibinfo {address} {Cambridge},\
  \bibinfo {year} {2008})\BibitemShut {NoStop}%
\bibitem [{\citenamefont {Muryshev}\ \emph {et~al.}(2002)\citenamefont
  {Muryshev}, \citenamefont {Shlyapnikov}, \citenamefont {Ertmer},
  \citenamefont {Sengstock},\ and\ \citenamefont
  {Lewenstein}}]{Muryshev_PRL02}%
  \BibitemOpen
  \bibfield  {author} {\bibinfo {author} {\bibfnamefont {A.}~\bibnamefont
  {Muryshev}}, \bibinfo {author} {\bibfnamefont {G.~V.}\ \bibnamefont
  {Shlyapnikov}}, \bibinfo {author} {\bibfnamefont {W.}~\bibnamefont {Ertmer}},
  \bibinfo {author} {\bibfnamefont {K.}~\bibnamefont {Sengstock}}, \ and\
  \bibinfo {author} {\bibfnamefont {M.}~\bibnamefont {Lewenstein}},\ }\href
  {\doibase 10.1103/PhysRevLett.89.110401} {\bibfield  {journal} {\bibinfo
  {journal} {Physical Review Letters}\ }\textbf {\bibinfo {volume} {89}},\
  \bibinfo {pages} {110401} (\bibinfo {year} {2002})}\BibitemShut {NoStop}%
\bibitem [{\citenamefont {Salasnich}\ \emph
  {et~al.}(2002{\natexlab{a}})\citenamefont {Salasnich}, \citenamefont
  {Parola},\ and\ \citenamefont {Reatto}}]{Salasnich_PRA02}%
  \BibitemOpen
  \bibfield  {author} {\bibinfo {author} {\bibfnamefont {L.}~\bibnamefont
  {Salasnich}}, \bibinfo {author} {\bibfnamefont {A.}~\bibnamefont {Parola}}, \
  and\ \bibinfo {author} {\bibfnamefont {L.}~\bibnamefont {Reatto}},\ }\href
  {\doibase 10.1103/PhysRevA.65.043614} {\bibfield  {journal} {\bibinfo
  {journal} {Physical Review A}\ }\textbf {\bibinfo {volume} {65}},\ \bibinfo
  {pages} {043614} (\bibinfo {year} {2002}{\natexlab{a}})}\BibitemShut
  {NoStop}%
\bibitem [{\citenamefont {Salasnich}\ \emph
  {et~al.}(2002{\natexlab{b}})\citenamefont {Salasnich}, \citenamefont
  {Parola},\ and\ \citenamefont {Reatto}}]{Salasnich_PRA02-2}%
  \BibitemOpen
  \bibfield  {author} {\bibinfo {author} {\bibfnamefont {L.}~\bibnamefont
  {Salasnich}}, \bibinfo {author} {\bibfnamefont {A.}~\bibnamefont {Parola}}, \
  and\ \bibinfo {author} {\bibfnamefont {L.}~\bibnamefont {Reatto}},\ }\href
  {\doibase 10.1103/PhysRevA.66.043603} {\bibfield  {journal} {\bibinfo
  {journal} {Physical Review A}\ }\textbf {\bibinfo {volume} {66}},\ \bibinfo
  {pages} {043603} (\bibinfo {year} {2002}{\natexlab{b}})}\BibitemShut
  {NoStop}%
\bibitem [{\citenamefont {Massignan}\ and\ \citenamefont
  {Modugno}(2003)}]{Massignan_PRA03}%
  \BibitemOpen
  \bibfield  {author} {\bibinfo {author} {\bibfnamefont {P.}~\bibnamefont
  {Massignan}}\ and\ \bibinfo {author} {\bibfnamefont {M.}~\bibnamefont
  {Modugno}},\ }\href {\doibase 10.1103/PhysRevA.67.023614} {\bibfield
  {journal} {\bibinfo  {journal} {Physical Review A}\ }\textbf {\bibinfo
  {volume} {67}},\ \bibinfo {pages} {023614} (\bibinfo {year}
  {2003})}\BibitemShut {NoStop}%
\bibitem [{\citenamefont {Kamchatnov}\ and\ \citenamefont
  {Shchesnovich}(2004)}]{Kamchatnov_PRA04}%
  \BibitemOpen
  \bibfield  {author} {\bibinfo {author} {\bibfnamefont {A.~M.}\ \bibnamefont
  {Kamchatnov}}\ and\ \bibinfo {author} {\bibfnamefont {V.~S.}\ \bibnamefont
  {Shchesnovich}},\ }\href {\doibase 10.1103/PhysRevA.70.023604} {\bibfield
  {journal} {\bibinfo  {journal} {Physical Review A}\ }\textbf {\bibinfo
  {volume} {70}},\ \bibinfo {pages} {023604} (\bibinfo {year}
  {2004})}\BibitemShut {NoStop}%
\bibitem [{\citenamefont {Carr}\ and\ \citenamefont
  {Brand}(2004)}]{Carr_PRL04}%
  \BibitemOpen
  \bibfield  {author} {\bibinfo {author} {\bibfnamefont {L.~D.}\ \bibnamefont
  {Carr}}\ and\ \bibinfo {author} {\bibfnamefont {J.}~\bibnamefont {Brand}},\
  }\href {\doibase 10.1103/PhysRevLett.92.040401} {\bibfield  {journal}
  {\bibinfo  {journal} {Physical Review Letters}\ }\textbf {\bibinfo {volume}
  {92}},\ \bibinfo {pages} {040401} (\bibinfo {year} {2004})}\BibitemShut
  {NoStop}%
\bibitem [{\citenamefont {Zhang}\ and\ \citenamefont
  {You}(2005)}]{Zhang_PRA05}%
  \BibitemOpen
  \bibfield  {author} {\bibinfo {author} {\bibfnamefont {W.}~\bibnamefont
  {Zhang}}\ and\ \bibinfo {author} {\bibfnamefont {L.}~\bibnamefont {You}},\
  }\href {\doibase 10.1103/PhysRevA.71.025603} {\bibfield  {journal} {\bibinfo
  {journal} {Physical Review A}\ }\textbf {\bibinfo {volume} {71}},\ \bibinfo
  {pages} {025603} (\bibinfo {year} {2005})}\BibitemShut {NoStop}%
\bibitem [{\citenamefont {Salasnich}\ and\ \citenamefont
  {Malomed}(2006)}]{Salasnich_PRA06}%
  \BibitemOpen
  \bibfield  {author} {\bibinfo {author} {\bibfnamefont {L.}~\bibnamefont
  {Salasnich}}\ and\ \bibinfo {author} {\bibfnamefont {B.~A.}\ \bibnamefont
  {Malomed}},\ }\href {\doibase 10.1103/PhysRevA.74.053610} {\bibfield
  {journal} {\bibinfo  {journal} {Physical Review A}\ }\textbf {\bibinfo
  {volume} {74}},\ \bibinfo {pages} {053610} (\bibinfo {year}
  {2006})}\BibitemShut {NoStop}%
\bibitem [{\citenamefont {{Matuszewski Micha{\l}and Kr{\'{o}}likowski}}\ \emph
  {et~al.}(2006)\citenamefont {{Matuszewski Micha{\l}and Kr{\'{o}}likowski}},
  \citenamefont {Trippenbach}, \citenamefont {Kivshar}, \citenamefont
  {Matuszewski}, \citenamefont {Kr{\'{o}}likowski}, \citenamefont
  {Trippenbach}, \citenamefont {Kivshar}, \citenamefont {{Matuszewski
  Micha{\l}and Kr{\'{o}}likowski}}, \citenamefont {Trippenbach},\ and\
  \citenamefont {Kivshar}}]{Matuszewski_PRA06}%
  \BibitemOpen
  \bibfield  {author} {\bibinfo {author} {\bibfnamefont {W.}~\bibnamefont
  {{Matuszewski Micha{\l}and Kr{\'{o}}likowski}}}, \bibinfo {author}
  {\bibfnamefont {M.}~\bibnamefont {Trippenbach}}, \bibinfo {author}
  {\bibfnamefont {Y.~S.}\ \bibnamefont {Kivshar}}, \bibinfo {author}
  {\bibfnamefont {M.}~\bibnamefont {Matuszewski}}, \bibinfo {author}
  {\bibfnamefont {W.}~\bibnamefont {Kr{\'{o}}likowski}}, \bibinfo {author}
  {\bibfnamefont {M.}~\bibnamefont {Trippenbach}}, \bibinfo {author}
  {\bibfnamefont {Y.~S.}\ \bibnamefont {Kivshar}}, \bibinfo {author}
  {\bibfnamefont {W.}~\bibnamefont {{Matuszewski Micha{\l}and
  Kr{\'{o}}likowski}}}, \bibinfo {author} {\bibfnamefont {M.}~\bibnamefont
  {Trippenbach}}, \ and\ \bibinfo {author} {\bibfnamefont {Y.~S.}\ \bibnamefont
  {Kivshar}},\ }\href {\doibase 10.1103/PhysRevA.73.063621} {\bibfield
  {journal} {\bibinfo  {journal} {Phys. Rev. A}\ }\textbf {\bibinfo {volume}
  {73}},\ \bibinfo {pages} {63621} (\bibinfo {year} {2006})}\BibitemShut
  {NoStop}%
\bibitem [{\citenamefont {Wei}\ \emph {et~al.}(2007)\citenamefont {Wei},
  \citenamefont {Zhi-Bing},\ and\ \citenamefont {Cheng-Guang}}]{Wei_CPL07}%
  \BibitemOpen
  \bibfield  {author} {\bibinfo {author} {\bibfnamefont {P.}~\bibnamefont
  {Wei}}, \bibinfo {author} {\bibfnamefont {L.}~\bibnamefont {Zhi-Bing}}, \
  and\ \bibinfo {author} {\bibfnamefont {B.}~\bibnamefont {Cheng-Guang}},\
  }\href {\doibase 10.1088/0256-307X/24/10/009} {\bibfield  {journal} {\bibinfo
   {journal} {Chinese Physics Letters}\ }\textbf {\bibinfo {volume} {24}},\
  \bibinfo {pages} {2745} (\bibinfo {year} {2007})}\BibitemShut {NoStop}%
\bibitem [{\citenamefont {Salasnich}\ \emph
  {et~al.}(2007{\natexlab{a}})\citenamefont {Salasnich}, \citenamefont
  {Cetoli}, \citenamefont {Malomed}, \citenamefont {Toigo},\ and\ \citenamefont
  {Reatto}}]{Salasnich_PRA07}%
  \BibitemOpen
  \bibfield  {author} {\bibinfo {author} {\bibfnamefont {L.}~\bibnamefont
  {Salasnich}}, \bibinfo {author} {\bibfnamefont {A.}~\bibnamefont {Cetoli}},
  \bibinfo {author} {\bibfnamefont {B.~A.}\ \bibnamefont {Malomed}}, \bibinfo
  {author} {\bibfnamefont {F.}~\bibnamefont {Toigo}}, \ and\ \bibinfo {author}
  {\bibfnamefont {L.}~\bibnamefont {Reatto}},\ }\href {\doibase
  10.1103/PhysRevA.76.013623} {\bibfield  {journal} {\bibinfo  {journal}
  {Physical Review A}\ }\textbf {\bibinfo {volume} {76}},\ \bibinfo {pages}
  {013623} (\bibinfo {year} {2007}{\natexlab{a}})}\BibitemShut {NoStop}%
\bibitem [{\citenamefont {Salasnich}\ \emph
  {et~al.}(2007{\natexlab{b}})\citenamefont {Salasnich}, \citenamefont
  {Cetoli}, \citenamefont {Malomed},\ and\ \citenamefont
  {Toigo}}]{Salasnich_PRA07-2}%
  \BibitemOpen
  \bibfield  {author} {\bibinfo {author} {\bibfnamefont {L.}~\bibnamefont
  {Salasnich}}, \bibinfo {author} {\bibfnamefont {A.}~\bibnamefont {Cetoli}},
  \bibinfo {author} {\bibfnamefont {B.~A.}\ \bibnamefont {Malomed}}, \ and\
  \bibinfo {author} {\bibfnamefont {F.}~\bibnamefont {Toigo}},\ }\href
  {\doibase 10.1103/PhysRevA.75.033622} {\bibfield  {journal} {\bibinfo
  {journal} {Physical Review A}\ }\textbf {\bibinfo {volume} {75}},\ \bibinfo
  {pages} {033622} (\bibinfo {year} {2007}{\natexlab{b}})}\BibitemShut
  {NoStop}%
\bibitem [{\citenamefont {Mateo}\ and\ \citenamefont
  {Delgado}(2007)}]{Mateo_PRA07}%
  \BibitemOpen
  \bibfield  {author} {\bibinfo {author} {\bibfnamefont {A.~M.}\ \bibnamefont
  {Mateo}}\ and\ \bibinfo {author} {\bibfnamefont {V.}~\bibnamefont
  {Delgado}},\ }\href {\doibase 10.1103/PhysRevA.75.063610} {\bibfield
  {journal} {\bibinfo  {journal} {Physical Review A}\ }\textbf {\bibinfo
  {volume} {75}},\ \bibinfo {pages} {063610} (\bibinfo {year}
  {2007})}\BibitemShut {NoStop}%
\bibitem [{\citenamefont {Maluckov}\ \emph {et~al.}(2008)\citenamefont
  {Maluckov}, \citenamefont {Had{\v{z}}ievski}, \citenamefont {Malomed},\ and\
  \citenamefont {Salasnich}}]{Maluckov_PRA08}%
  \BibitemOpen
  \bibfield  {author} {\bibinfo {author} {\bibfnamefont {A.}~\bibnamefont
  {Maluckov}}, \bibinfo {author} {\bibfnamefont {L.}~\bibnamefont
  {Had{\v{z}}ievski}}, \bibinfo {author} {\bibfnamefont {B.~A.}\ \bibnamefont
  {Malomed}}, \ and\ \bibinfo {author} {\bibfnamefont {L.}~\bibnamefont
  {Salasnich}},\ }\href {\doibase 10.1103/PhysRevA.78.013616} {\bibfield
  {journal} {\bibinfo  {journal} {Physical Review A}\ }\textbf {\bibinfo
  {volume} {78}},\ \bibinfo {pages} {013616} (\bibinfo {year}
  {2008})}\BibitemShut {NoStop}%
\bibitem [{\citenamefont {Salasnich}\ \emph {et~al.}(2008)\citenamefont
  {Salasnich}, \citenamefont {Malomed},\ and\ \citenamefont
  {Toigo}}]{Salasnich_PRA08}%
  \BibitemOpen
  \bibfield  {author} {\bibinfo {author} {\bibfnamefont {L.}~\bibnamefont
  {Salasnich}}, \bibinfo {author} {\bibfnamefont {B.~A.}\ \bibnamefont
  {Malomed}}, \ and\ \bibinfo {author} {\bibfnamefont {F.}~\bibnamefont
  {Toigo}},\ }\href {\doibase 10.1103/PhysRevA.77.035601} {\bibfield  {journal}
  {\bibinfo  {journal} {Physical Review A}\ }\textbf {\bibinfo {volume} {77}},\
  \bibinfo {pages} {035601} (\bibinfo {year} {2008})}\BibitemShut {NoStop}%
\bibitem [{\citenamefont {Salasnich}(2009)}]{Salasnich_JPA09}%
  \BibitemOpen
  \bibfield  {author} {\bibinfo {author} {\bibfnamefont {L.}~\bibnamefont
  {Salasnich}},\ }\href {\doibase 10.1088/1751-8113/42/33/335205} {\bibfield
  {journal} {\bibinfo  {journal} {Journal of Physics A: Mathematical and
  Theoretical}\ }\textbf {\bibinfo {volume} {42}},\ \bibinfo {pages} {335205}
  (\bibinfo {year} {2009})}\BibitemShut {NoStop}%
\bibitem [{\citenamefont {Li}\ \emph {et~al.}(2009)\citenamefont {Li},
  \citenamefont {Guang-Yuan}, \citenamefont {Yong-Jun}, \citenamefont
  {Xian-Feng},\ and\ \citenamefont {Jiu-Rong}}]{Li_CTP09}%
  \BibitemOpen
  \bibfield  {author} {\bibinfo {author} {\bibfnamefont {Y.}~\bibnamefont
  {Li}}, \bibinfo {author} {\bibfnamefont {X.}~\bibnamefont {Guang-Yuan}},
  \bibinfo {author} {\bibfnamefont {W.}~\bibnamefont {Yong-Jun}}, \bibinfo
  {author} {\bibfnamefont {L.}~\bibnamefont {Xian-Feng}}, \ and\ \bibinfo
  {author} {\bibfnamefont {H.}~\bibnamefont {Jiu-Rong}},\ }\href {\doibase
  10.1088/0253-6102/52/3/10} {\bibfield  {journal} {\bibinfo  {journal}
  {Communications in Theoretical Physics}\ }\textbf {\bibinfo {volume} {52}},\
  \bibinfo {pages} {431} (\bibinfo {year} {2009})}\BibitemShut {NoStop}%
\bibitem [{\citenamefont {Buitrago}\ and\ \citenamefont
  {Adhikari}(2009)}]{Buitrago_JPB09}%
  \BibitemOpen
  \bibfield  {author} {\bibinfo {author} {\bibfnamefont {C.~A.~G.}\
  \bibnamefont {Buitrago}}\ and\ \bibinfo {author} {\bibfnamefont {S.~K.}\
  \bibnamefont {Adhikari}},\ }\href {\doibase 10.1088/0953-4075/42/21/215306}
  {\bibfield  {journal} {\bibinfo  {journal} {Journal of Physics B: Atomic,
  Molecular and Optical Physics}\ }\textbf {\bibinfo {volume} {42}},\ \bibinfo
  {pages} {215306} (\bibinfo {year} {2009})}\BibitemShut {NoStop}%
\bibitem [{\citenamefont {Adhikari}\ and\ \citenamefont
  {Salasnich}(2009)}]{Adhikari_NJP09}%
  \BibitemOpen
  \bibfield  {author} {\bibinfo {author} {\bibfnamefont {S.~K.}\ \bibnamefont
  {Adhikari}}\ and\ \bibinfo {author} {\bibfnamefont {L.}~\bibnamefont
  {Salasnich}},\ }\href {\doibase 10.1088/1367-2630/11/2/023011} {\bibfield
  {journal} {\bibinfo  {journal} {New Journal of Physics}\ }\textbf {\bibinfo
  {volume} {11}},\ \bibinfo {pages} {023011} (\bibinfo {year}
  {2009})}\BibitemShut {NoStop}%
\bibitem [{\citenamefont {Salasnich}\ and\ \citenamefont
  {Malomed}(2009)}]{Salasnich_PRA09}%
  \BibitemOpen
  \bibfield  {author} {\bibinfo {author} {\bibfnamefont {L.}~\bibnamefont
  {Salasnich}}\ and\ \bibinfo {author} {\bibfnamefont {B.~A.}\ \bibnamefont
  {Malomed}},\ }\href {\doibase 10.1103/PhysRevA.79.053620} {\bibfield
  {journal} {\bibinfo  {journal} {Physical Review A}\ }\textbf {\bibinfo
  {volume} {79}},\ \bibinfo {pages} {053620} (\bibinfo {year}
  {2009})}\BibitemShut {NoStop}%
\bibitem [{\citenamefont {{Mu{\~{n}}oz Mateo}}\ and\ \citenamefont
  {Delgado}(2009)}]{Munoz-Mateo_AP09}%
  \BibitemOpen
  \bibfield  {author} {\bibinfo {author} {\bibfnamefont {A.}~\bibnamefont
  {{Mu{\~{n}}oz Mateo}}}\ and\ \bibinfo {author} {\bibfnamefont
  {V.}~\bibnamefont {Delgado}},\ }\href {\doibase 10.1016/j.aop.2008.10.002}
  {\bibfield  {journal} {\bibinfo  {journal} {Annals of Physics}\ }\textbf
  {\bibinfo {volume} {324}},\ \bibinfo {pages} {709} (\bibinfo {year}
  {2009})}\BibitemShut {NoStop}%
\bibitem [{\citenamefont {Young-S.}\ \emph {et~al.}(2010)\citenamefont
  {Young-S.}, \citenamefont {Salasnich},\ and\ \citenamefont
  {Adhikari}}]{Young_PRA10}%
  \BibitemOpen
  \bibfield  {author} {\bibinfo {author} {\bibfnamefont {L.~E.}\ \bibnamefont
  {Young-S.}}, \bibinfo {author} {\bibfnamefont {L.}~\bibnamefont {Salasnich}},
  \ and\ \bibinfo {author} {\bibfnamefont {S.~K.}\ \bibnamefont {Adhikari}},\
  }\href {\doibase 10.1103/PhysRevA.82.053601} {\bibfield  {journal} {\bibinfo
  {journal} {Physical Review A}\ }\textbf {\bibinfo {volume} {82}},\ \bibinfo
  {pages} {053601} (\bibinfo {year} {2010})}\BibitemShut {NoStop}%
\bibitem [{\citenamefont {Cardoso}\ \emph {et~al.}(2011)\citenamefont
  {Cardoso}, \citenamefont {Avelar},\ and\ \citenamefont
  {Bazeia}}]{Cardoso_PRE11}%
  \BibitemOpen
  \bibfield  {author} {\bibinfo {author} {\bibfnamefont {W.~B.}\ \bibnamefont
  {Cardoso}}, \bibinfo {author} {\bibfnamefont {A.~T.}\ \bibnamefont {Avelar}},
  \ and\ \bibinfo {author} {\bibfnamefont {D.}~\bibnamefont {Bazeia}},\ }\href
  {\doibase 10.1103/PhysRevE.83.036604} {\bibfield  {journal} {\bibinfo
  {journal} {Physical Review E}\ }\textbf {\bibinfo {volume} {83}},\ \bibinfo
  {pages} {036604} (\bibinfo {year} {2011})}\BibitemShut {NoStop}%
\bibitem [{\citenamefont {Mateo}\ \emph {et~al.}(2011)\citenamefont {Mateo},
  \citenamefont {Delgado},\ and\ \citenamefont {Malomed}}]{Mateo_PRA11}%
  \BibitemOpen
  \bibfield  {author} {\bibinfo {author} {\bibfnamefont {A.~M.}\ \bibnamefont
  {Mateo}}, \bibinfo {author} {\bibfnamefont {V.}~\bibnamefont {Delgado}}, \
  and\ \bibinfo {author} {\bibfnamefont {B.~A.}\ \bibnamefont {Malomed}},\
  }\href {\doibase 10.1103/PhysRevA.83.053610} {\bibfield  {journal} {\bibinfo
  {journal} {Physical Review A}\ }\textbf {\bibinfo {volume} {83}},\ \bibinfo
  {pages} {053610} (\bibinfo {year} {2011})}\BibitemShut {NoStop}%
\bibitem [{\citenamefont {Edwards}\ \emph {et~al.}(2012)\citenamefont
  {Edwards}, \citenamefont {Krygier}, \citenamefont {Seddiqi}, \citenamefont
  {Benton},\ and\ \citenamefont {Clark}}]{Edwards_PRE12}%
  \BibitemOpen
  \bibfield  {author} {\bibinfo {author} {\bibfnamefont {M.}~\bibnamefont
  {Edwards}}, \bibinfo {author} {\bibfnamefont {M.}~\bibnamefont {Krygier}},
  \bibinfo {author} {\bibfnamefont {H.}~\bibnamefont {Seddiqi}}, \bibinfo
  {author} {\bibfnamefont {B.}~\bibnamefont {Benton}}, \ and\ \bibinfo {author}
  {\bibfnamefont {C.~W.}\ \bibnamefont {Clark}},\ }\href {\doibase
  10.1103/PhysRevE.86.056710} {\bibfield  {journal} {\bibinfo  {journal}
  {Physical Review E}\ }\textbf {\bibinfo {volume} {86}},\ \bibinfo {pages}
  {056710} (\bibinfo {year} {2012})}\BibitemShut {NoStop}%
\bibitem [{\citenamefont {D{\'{i}}az}\ \emph {et~al.}(2012)\citenamefont
  {D{\'{i}}az}, \citenamefont {Laroze}, \citenamefont {Schmidt},\ and\
  \citenamefont {Malomed}}]{Diaz_JPB12}%
  \BibitemOpen
  \bibfield  {author} {\bibinfo {author} {\bibfnamefont {P.}~\bibnamefont
  {D{\'{i}}az}}, \bibinfo {author} {\bibfnamefont {D.}~\bibnamefont {Laroze}},
  \bibinfo {author} {\bibfnamefont {I.}~\bibnamefont {Schmidt}}, \ and\
  \bibinfo {author} {\bibfnamefont {B.~A.}\ \bibnamefont {Malomed}},\ }\href
  {\doibase 10.1088/0953-4075/45/14/145304} {\bibfield  {journal} {\bibinfo
  {journal} {Journal of Physics B: Atomic, Molecular and Optical Physics}\
  }\textbf {\bibinfo {volume} {45}},\ \bibinfo {pages} {145304} (\bibinfo
  {year} {2012})}\BibitemShut {NoStop}%
\bibitem [{\citenamefont {Salasnich}\ and\ \citenamefont
  {Malomed}(2013)}]{Salasnich_PRA13}%
  \BibitemOpen
  \bibfield  {author} {\bibinfo {author} {\bibfnamefont {L.}~\bibnamefont
  {Salasnich}}\ and\ \bibinfo {author} {\bibfnamefont {B.~A.}\ \bibnamefont
  {Malomed}},\ }\href {\doibase 10.1103/PhysRevA.87.063625} {\bibfield
  {journal} {\bibinfo  {journal} {Physical Review A}\ }\textbf {\bibinfo
  {volume} {87}},\ \bibinfo {pages} {063625} (\bibinfo {year}
  {2013})}\BibitemShut {NoStop}%
\bibitem [{\citenamefont {Salasnich}\ \emph {et~al.}(2014)\citenamefont
  {Salasnich}, \citenamefont {Cardoso},\ and\ \citenamefont
  {Malomed}}]{Salasnich_PRA14}%
  \BibitemOpen
  \bibfield  {author} {\bibinfo {author} {\bibfnamefont {L.}~\bibnamefont
  {Salasnich}}, \bibinfo {author} {\bibfnamefont {W.~B.}\ \bibnamefont
  {Cardoso}}, \ and\ \bibinfo {author} {\bibfnamefont {B.~A.}\ \bibnamefont
  {Malomed}},\ }\href {\doibase 10.1103/PhysRevA.90.033629} {\bibfield
  {journal} {\bibinfo  {journal} {Physical Review A}\ }\textbf {\bibinfo
  {volume} {90}},\ \bibinfo {pages} {033629} (\bibinfo {year}
  {2014})}\BibitemShut {NoStop}%
\bibitem [{\citenamefont {Chiquillo}(2015)}]{Chiquillo_JPA15}%
  \BibitemOpen
  \bibfield  {author} {\bibinfo {author} {\bibfnamefont {E.}~\bibnamefont
  {Chiquillo}},\ }\href {\doibase 10.1088/1751-8113/48/47/475001} {\bibfield
  {journal} {\bibinfo  {journal} {Journal of Physics A: Mathematical and
  Theoretical}\ }\textbf {\bibinfo {volume} {48}},\ \bibinfo {pages} {475001}
  (\bibinfo {year} {2015})}\BibitemShut {NoStop}%
\bibitem [{\citenamefont {dos Santos}\ and\ \citenamefont
  {Cardoso}(2019)}]{Calixto1}%
  \BibitemOpen
  \bibfield  {author} {\bibinfo {author} {\bibfnamefont {M.~C.~P.}\
  \bibnamefont {dos Santos}}\ and\ \bibinfo {author} {\bibfnamefont {W.~B.}\
  \bibnamefont {Cardoso}},\ }\href {\doibase 10.1016/j.physleta.2019.01.064}
  {\bibfield  {journal} {\bibinfo  {journal} {Phys. Lett. A}\ }\textbf
  {\bibinfo {volume} {383}},\ \bibinfo {pages} {250401} (\bibinfo {year}
  {2019})}\BibitemShut {NoStop}%
\bibitem [{\citenamefont {Cuevas}\ \emph {et~al.}(2013)\citenamefont {Cuevas},
  \citenamefont {Kevrekidis}, \citenamefont {Malomed}, \citenamefont {Dyke},\
  and\ \citenamefont {Hulet}}]{Cuevas_NJP13}%
  \BibitemOpen
  \bibfield  {author} {\bibinfo {author} {\bibfnamefont {J.}~\bibnamefont
  {Cuevas}}, \bibinfo {author} {\bibfnamefont {P.~G.}\ \bibnamefont
  {Kevrekidis}}, \bibinfo {author} {\bibfnamefont {B.~A.}\ \bibnamefont
  {Malomed}}, \bibinfo {author} {\bibfnamefont {P.}~\bibnamefont {Dyke}}, \
  and\ \bibinfo {author} {\bibfnamefont {R.~G.}\ \bibnamefont {Hulet}},\ }\href
  {\doibase 10.1088/1367-2630/15/6/063006} {\bibfield  {journal} {\bibinfo
  {journal} {New Journal of Physics}\ }\textbf {\bibinfo {volume} {15}},\
  \bibinfo {pages} {063006} (\bibinfo {year} {2013})}\BibitemShut {NoStop}%
\bibitem [{\citenamefont {{De Nicola}}\ \emph {et~al.}(2006)\citenamefont {{De
  Nicola}}, \citenamefont {Malomed},\ and\ \citenamefont
  {Fedele}}]{DeNicola_PLA06}%
  \BibitemOpen
  \bibfield  {author} {\bibinfo {author} {\bibfnamefont {S.}~\bibnamefont {{De
  Nicola}}}, \bibinfo {author} {\bibfnamefont {B.~A.}\ \bibnamefont {Malomed}},
  \ and\ \bibinfo {author} {\bibfnamefont {R.}~\bibnamefont {Fedele}},\ }\href
  {\doibase 10.1016/j.physleta.2006.07.062} {\bibfield  {journal} {\bibinfo
  {journal} {Physics Letters A}\ }\textbf {\bibinfo {volume} {360}},\ \bibinfo
  {pages} {164} (\bibinfo {year} {2006})}\BibitemShut {NoStop}%
\bibitem [{\citenamefont {Li}\ \emph {et~al.}(2017)\citenamefont {Li},
  \citenamefont {Luo}, \citenamefont {Liu}, \citenamefont {Chen}, \citenamefont
  {Huang}, \citenamefont {Fu}, \citenamefont {Tan},\ and\ \citenamefont
  {Malomed}}]{Li_NJP17}%
  \BibitemOpen
  \bibfield  {author} {\bibinfo {author} {\bibfnamefont {Y.}~\bibnamefont
  {Li}}, \bibinfo {author} {\bibfnamefont {Z.}~\bibnamefont {Luo}}, \bibinfo
  {author} {\bibfnamefont {Y.}~\bibnamefont {Liu}}, \bibinfo {author}
  {\bibfnamefont {Z.}~\bibnamefont {Chen}}, \bibinfo {author} {\bibfnamefont
  {C.}~\bibnamefont {Huang}}, \bibinfo {author} {\bibfnamefont
  {S.}~\bibnamefont {Fu}}, \bibinfo {author} {\bibfnamefont {H.}~\bibnamefont
  {Tan}}, \ and\ \bibinfo {author} {\bibfnamefont {B.~A.}\ \bibnamefont
  {Malomed}},\ }\href {\doibase 10.1088/1367-2630/aa983b} {\bibfield  {journal}
  {\bibinfo  {journal} {New Journal of Physics}\ }\textbf {\bibinfo {volume}
  {19}},\ \bibinfo {pages} {113043} (\bibinfo {year} {2017})}\BibitemShut
  {NoStop}%
\bibitem [{\citenamefont {Li}\ \emph {et~al.}(2018)\citenamefont {Li},
  \citenamefont {Chen}, \citenamefont {Luo}, \citenamefont {Huang},
  \citenamefont {Tan}, \citenamefont {Pang},\ and\ \citenamefont
  {Malomed}}]{Li_PRA18}%
  \BibitemOpen
  \bibfield  {author} {\bibinfo {author} {\bibfnamefont {Y.}~\bibnamefont
  {Li}}, \bibinfo {author} {\bibfnamefont {Z.}~\bibnamefont {Chen}}, \bibinfo
  {author} {\bibfnamefont {Z.}~\bibnamefont {Luo}}, \bibinfo {author}
  {\bibfnamefont {C.}~\bibnamefont {Huang}}, \bibinfo {author} {\bibfnamefont
  {H.}~\bibnamefont {Tan}}, \bibinfo {author} {\bibfnamefont {W.}~\bibnamefont
  {Pang}}, \ and\ \bibinfo {author} {\bibfnamefont {B.~A.}\ \bibnamefont
  {Malomed}},\ }\href {\doibase 10.1103/PhysRevA.98.063602} {\bibfield
  {journal} {\bibinfo  {journal} {Physical Review A}\ }\textbf {\bibinfo
  {volume} {98}},\ \bibinfo {pages} {063602} (\bibinfo {year}
  {2018})}\BibitemShut {NoStop}%
\bibitem [{\citenamefont {Sakaguchi}\ and\ \citenamefont
  {Malomed}(2011)}]{Sakaguchi_PRA11}%
  \BibitemOpen
  \bibfield  {author} {\bibinfo {author} {\bibfnamefont {H.}~\bibnamefont
  {Sakaguchi}}\ and\ \bibinfo {author} {\bibfnamefont {B.~A.}\ \bibnamefont
  {Malomed}},\ }\href {\doibase 10.1103/PhysRevA.83.013607} {\bibfield
  {journal} {\bibinfo  {journal} {Physical Review A}\ }\textbf {\bibinfo
  {volume} {83}},\ \bibinfo {pages} {013607} (\bibinfo {year}
  {2011})}\BibitemShut {NoStop}%
\bibitem [{\citenamefont {Denschlag}\ and\ \citenamefont
  {Schmiedmayer}(1997)}]{Denschlag_EPL97}%
  \BibitemOpen
  \bibfield  {author} {\bibinfo {author} {\bibfnamefont {J.}~\bibnamefont
  {Denschlag}}\ and\ \bibinfo {author} {\bibfnamefont {J.}~\bibnamefont
  {Schmiedmayer}},\ }\href {\doibase 10.1209/epl/i1997-00259-y} {\bibfield
  {journal} {\bibinfo  {journal} {Europhysics Letters (EPL)}\ }\textbf
  {\bibinfo {volume} {38}},\ \bibinfo {pages} {405} (\bibinfo {year}
  {1997})}\BibitemShut {NoStop}%
\bibitem [{\citenamefont {Ferrier-Barbut}\ \emph {et~al.}(2016)\citenamefont
  {Ferrier-Barbut}, \citenamefont {Kadau}, \citenamefont {Schmitt},
  \citenamefont {Wenzel},\ and\ \citenamefont {Pfau}}]{Ferrier-Barbut_PRL16}%
  \BibitemOpen
  \bibfield  {author} {\bibinfo {author} {\bibfnamefont {I.}~\bibnamefont
  {Ferrier-Barbut}}, \bibinfo {author} {\bibfnamefont {H.}~\bibnamefont
  {Kadau}}, \bibinfo {author} {\bibfnamefont {M.}~\bibnamefont {Schmitt}},
  \bibinfo {author} {\bibfnamefont {M.}~\bibnamefont {Wenzel}}, \ and\ \bibinfo
  {author} {\bibfnamefont {T.}~\bibnamefont {Pfau}},\ }\href {\doibase
  10.1103/PhysRevLett.116.215301} {\bibfield  {journal} {\bibinfo  {journal}
  {Physical Review Letters}\ }\textbf {\bibinfo {volume} {116}},\ \bibinfo
  {pages} {215301} (\bibinfo {year} {2016})}\BibitemShut {NoStop}%
\bibitem [{\citenamefont {Schmitt}\ \emph {et~al.}(2016)\citenamefont
  {Schmitt}, \citenamefont {Wenzel}, \citenamefont {B{\"{o}}ttcher},
  \citenamefont {Ferrier-Barbut},\ and\ \citenamefont {Pfau}}]{Schmitt_NAT16}%
  \BibitemOpen
  \bibfield  {author} {\bibinfo {author} {\bibfnamefont {M.}~\bibnamefont
  {Schmitt}}, \bibinfo {author} {\bibfnamefont {M.}~\bibnamefont {Wenzel}},
  \bibinfo {author} {\bibfnamefont {F.}~\bibnamefont {B{\"{o}}ttcher}},
  \bibinfo {author} {\bibfnamefont {I.}~\bibnamefont {Ferrier-Barbut}}, \ and\
  \bibinfo {author} {\bibfnamefont {T.}~\bibnamefont {Pfau}},\ }\href {\doibase
  10.1038/nature20126} {\bibfield  {journal} {\bibinfo  {journal} {Nature}\
  }\textbf {\bibinfo {volume} {539}},\ \bibinfo {pages} {259} (\bibinfo {year}
  {2016})}\BibitemShut {NoStop}%
\bibitem [{\citenamefont {Chomaz}\ \emph {et~al.}(2016)\citenamefont {Chomaz},
  \citenamefont {Baier}, \citenamefont {Petter}, \citenamefont {Mark},
  \citenamefont {W{\"{a}}chtler}, \citenamefont {Santos},\ and\ \citenamefont
  {Ferlaino}}]{Chomaz_PRX16}%
  \BibitemOpen
  \bibfield  {author} {\bibinfo {author} {\bibfnamefont {L.}~\bibnamefont
  {Chomaz}}, \bibinfo {author} {\bibfnamefont {S.}~\bibnamefont {Baier}},
  \bibinfo {author} {\bibfnamefont {D.}~\bibnamefont {Petter}}, \bibinfo
  {author} {\bibfnamefont {M.~J.}\ \bibnamefont {Mark}}, \bibinfo {author}
  {\bibfnamefont {F.}~\bibnamefont {W{\"{a}}chtler}}, \bibinfo {author}
  {\bibfnamefont {L.}~\bibnamefont {Santos}}, \ and\ \bibinfo {author}
  {\bibfnamefont {F.}~\bibnamefont {Ferlaino}},\ }\href {\doibase
  10.1103/PhysRevX.6.041039} {\bibfield  {journal} {\bibinfo  {journal}
  {Physical Review X}\ }\textbf {\bibinfo {volume} {6}},\ \bibinfo {pages}
  {041039} (\bibinfo {year} {2016})}\BibitemShut {NoStop}%
\bibitem [{\citenamefont {Fibich}(2015)}]{Fibich_15}%
  \BibitemOpen
  \bibfield  {author} {\bibinfo {author} {\bibfnamefont {G.}~\bibnamefont
  {Fibich}},\ }\href {\doibase 10.1007/978-3-319-12748-4} {\emph {\bibinfo
  {title} {{The Nonlinear Schr{\"{o}}dinger Equation: Singular Solutions and
  Optical Collapse}}}},\ \bibinfo {series} {Applied Mathematical Sciences},
  Vol.\ \bibinfo {volume} {192}\ (\bibinfo  {publisher} {Springer International
  Publishing},\ \bibinfo {address} {Cham},\ \bibinfo {year} {2015})\BibitemShut
  {NoStop}%
\bibitem [{\citenamefont {Dziarmaga}(2010)}]{Dziarmaga_AP10}%
  \BibitemOpen
  \bibfield  {author} {\bibinfo {author} {\bibfnamefont {J.}~\bibnamefont
  {Dziarmaga}},\ }\href {\doibase 10.1080/00018732.2010.514702} {\bibfield
  {journal} {\bibinfo  {journal} {Advances in Physics}\ }\textbf {\bibinfo
  {volume} {59}},\ \bibinfo {pages} {1063} (\bibinfo {year}
  {2010})}\BibitemShut {NoStop}%
\bibitem [{\citenamefont {Krolikowski}\ and\ \citenamefont
  {Luther-Davies}(1993)}]{Krolikowski_OL93}%
  \BibitemOpen
  \bibfield  {author} {\bibinfo {author} {\bibfnamefont {W.}~\bibnamefont
  {Krolikowski}}\ and\ \bibinfo {author} {\bibfnamefont {B.}~\bibnamefont
  {Luther-Davies}},\ }\href {\doibase 10.1364/OL.18.000188} {\bibfield
  {journal} {\bibinfo  {journal} {Optics Letters}\ }\textbf {\bibinfo {volume}
  {18}},\ \bibinfo {pages} {188} (\bibinfo {year} {1993})}\BibitemShut
  {NoStop}%
\bibitem [{\citenamefont {Salasnich}\ \emph
  {et~al.}(2007{\natexlab{c}})\citenamefont {Salasnich}, \citenamefont
  {Malomed},\ and\ \citenamefont {Toigo}}]{Salasnich_PRA07.063614}%
  \BibitemOpen
  \bibfield  {author} {\bibinfo {author} {\bibfnamefont {L.}~\bibnamefont
  {Salasnich}}, \bibinfo {author} {\bibfnamefont {B.~A.}\ \bibnamefont
  {Malomed}}, \ and\ \bibinfo {author} {\bibfnamefont {F.}~\bibnamefont
  {Toigo}},\ }\href {\doibase 10.1103/PhysRevA.76.063614} {\bibfield  {journal}
  {\bibinfo  {journal} {Physical Review A}\ }\textbf {\bibinfo {volume} {76}},\
  \bibinfo {pages} {063614} (\bibinfo {year} {2007}{\natexlab{c}})}\BibitemShut
  {NoStop}%
\end{thebibliography}

%

\end{document}